\documentclass[xcolor=table, 12pt]{article}
\usepackage{amsmath, amsfonts, graphicx, graphics, bm, float, adjustbox, lscape, threeparttable, rotating, makecell, setspace, tabularx, multirow, booktabs, natbib, caption, subcaption, chngcntr, dsfont, url, enumitem}
\usepackage[table,xcdraw]{xcolor}
\usepackage[pdfauthor={},pdftitle={},pdfcreator={},pdfproducer={}]{}
\usepackage[utf8]{inputenc}
\usepackage[top=3cm,left=3cm,right=3cm,bottom=3cm]{geometry}
\usepackage[graphicx]{realboxes}
\usepackage [english]{babel}
\usepackage [autostyle, english = american]{csquotes}
\usepackage[all,cmtip]{xy} 
\MakeOuterQuote{"}
\usepackage[toc,page]{appendix}
\usepackage[T1]{fontenc}
\usepackage{hyperref}
\hypersetup{
colorlinks=true,
citecolor=blue,
linkcolor=blue,
filecolor=blue,      
urlcolor=blue,
}
\usepackage{authblk}
\usepackage{xurl}

\begin{document}

\begin{titlepage}
\title{Quantifying walkable accessibility to urban services: \\ An application to Florence, Italy}
\author[1]{Leonardo Boncinelli}
\author[1]{Stefania Miricola}
\author[2]{Eugenio Vicario}

\affil[1]{Department of Economics and Business, University of Florence, Via delle Pandette 9, 50127 Firenze, Italy}
\affil[2]{IMT School for Advanced Studies Lucca, Laboratory for the Analysis of compleX Economic Systems, Piazza S.~Francesco 19, 
            Lucca,
            55100, 
            Italy}

 \date{\today}
\maketitle
\vspace{0.1in}
\begin{abstract}
\footnotesize
\noindent  
The concept of quality of life in urban settings is increasingly associated to the accessibility of amenities within a short walking distance for residents. However, this narrative still requires thorough empirical investigation to evaluate the practical implications, benefits, and challenges. In this work, we propose a novel methodology for evaluating urban accessibility to services, with an application to the city of Florence, Italy. Our approach involves identifying the accessibility of essential services from residential buildings within a 10-minute walking distance, employing a rigorous spatial analysis process and open-source geospatial data. As a second contribution, we extend the concept of 10-minute accessibility within a network theory framework and apply a clustering algorithm to identify urban communities based on shared access to essential services. Finally, we explore the dimension of functional redundancy. Our proposed metrics represent a step forward towards an accurate assessment of the adherence to the 10-minute city model and offer a valuable tool for place-based policies aimed at addressing spatial disparities in urban development. \\

\vspace{0cm}
\vspace{3mm}
\bigskip
\end{abstract}
\setcounter{page}{0}
\thispagestyle{empty}
\end{titlepage}
\pagebreak \newpage

\onehalfspacing

\maketitle

\section{Introduction}
\label{sec: intro}

Designing cities to improve residents' quality of life is on the agenda of policymakers and institutions worldwide. This is evidenced by the establishment of guidelines to help urban planners prioritize quality of life in city design \citep{oecd2020cities} and the efforts in assessing well-being outcomes across cities \citep{oecd2022life}. But what does quality of life actually mean in urban contexts? In the economic field, it is traditionally understood as the direct utility a citizen gains from local consumption amenities, or, equivalently, as the appeal of a specific area as a place to live, independent of expected wages and cost-of-living factors \citep{roback1982wages}. Indeed, amenities play a key role in people's choice of where to live and cities' population density is positively associated with higher quality of life, i.e with the utility derived from local consumption amenities \citep{rappaport2008consumption}. Amenities are important also because they help explain what is observed in reality within cities, such as social stratification and varying degrees of social mix across neighborhoods. To better understand where people choose to live, models should go beyond just costs and income and include other factors like housing quality, neighbors, and local amenities. Considering these factors helps explain why urban areas often show social sorting, as highlighted in \cite{epple1981implications}, \cite{epple1998equilibrium}, and \cite{bayer2012tiebout}. \\


\noindent Current debates on this topic, informed by interdisciplinary perspectives, increasingly highlight the importance of spatial accessibility. This focus translates into a concrete emphasis on ensuring that essential services and amenities are within easy reach, whether by foot or public transport. This approach claims for a shift in urban design towards developing urban areas with autonomous and self-sufficient neighborhoods, moving away from the idea of a single city center as the focal point for all activities. This model gained significant attention, especially during the COVID-19 pandemic, which has reduced the value of city centers \citep{ROSENTHAL2022103381} and underscored the vulnerability of areas with poor access to services. A work in particular has contributed the most to popularizing the theme of spatial accessibility: in 2021 \cite{moreno2021introducing}'s introduced the concept of the \textit{15-minute city}, which advocates for a urban environment where citizens can reach key services, such as healthcare, education, and shopping, within a 15-minute walk or bike ride. \\


In this study, we propose a novel methodology for evaluating urban accessibility to services, with an application to the city of Florence, Italy. We assume that a citizen does have access to a specific service if the latter is within a 10-minute walking distance from its home. The analysis identifies accessible destinations from residential buildings, spanning six service macro-categories: Schools, Healthcare, Food Retail, Green Areas, Leisure, and Primary Services. We accordingly define a \textit{10-minute city Index}, which, as may have been noticed, lowers the time threshold required to consider a service accessible compared to the benchmark set by the popular 15-minute city concept. The choice is driven by the methodological innovations introduced in our approach: the use of buildings as the unit of analysis. This allows for precise and realistic representation of pedestrian route origins and enhances the granularity of urban accessibility measures to unprecedented levels. \\

Indeed, a plethora of studies have sought to operationalize \cite{moreno2021introducing}'s concept of a \textit{15-minute city}. However, there is no consensus on a unified approach and existing methodologies often show limitations. Previous studies utilizing mapping platform tools for travel distance calculations rely on starting points that only roughly approximate the actual spatial origin of individuals' daily commutes. To mention few of them, \citet{olivari2023italian} and \citet{murgante2024developing} use road intersections, \citet{staricco202215} and \citet{abbiasov202415} adopt census tract centroids, while \citet{akrami2024walk} and \citet{bruno2024universal} rely on grid centroids. By using buildings as the basis for isochrone calculations, our methodology achieves a higher spatial resolution that better reflects real-world conditions, providing a robust framework for assessing pedestrian accessibility in urban areas. \\

The precision achieved through the use of buildings as unit of analysis is pivotal as this index operates at a detailed spatial scale and is sensitive to marginal spatial variations. In this regard, this work adds to recent contributions in spatial analysis using buildings to detect human activities (see \cite{de2021delineating} using building density to identify urban areas or \cite{BILLINGS201613} using plant density to identify industrial agglomerations within urban areas). These methodologies using building-level data generate continuous measures that help overcome the well-known modifiable areal unit problem (MAUP), which arises when different ways of aggregating spatial data, by altering the shape or scale of regions, can significantly impact analytical results \citep{coombes1982use}. \\

This work is also related to urban studies on the compact city paradigm, discussing its advantages and drawbacks \citep{glaeser2004sprawl}. Indeed, the compact city closely parallels the 15-minute city model. Generally speaking, a compact city is characterized by a high residential density that is continuous throughout the urban area and is often presented as a sustainable alternative to urban sprawl. The latter is instead defined by the European Environment Agency (EEA) as the low-density expansion of large urban areas at the expense of agricultural land, characterized by mixed land use and suburban development \citep{EEA2006}. Urban sprawl has been criticized for its environmental impact, as it fragments the city into scarcely populated and disconnected areas that lack internal diversity, leading to increased car use and higher public service costs \citep{oecd2012compact}. Indeed, 
compact and dense urban areas have been proved to produce lower levels of $CO_2$ \citep{glaeser2010greenness}, contributing positively to the environment. \\





After presenting the \textit{10-minute Index}, we illustrate an application to the city of Florence.
We expect Florence to exhibit a good level of accessibility due to its compact urban layout and the relatively uniform distribution of its population. This pattern is indeed reflected in the data, with the \textit{10-minute Index} showing a heavily skewed right tail. Most residential buildings exhibit high accessibility values, mainly clustering in the upper quantiles of the distribution. Conversely, a review of the sub-indices shows that only a small portion of people have optimal access to green spaces and leisure services like sports and cultural activities. Notably, our index is designed to be universally applicable across urban contexts and facilitates comparisons between cities, providing valuable insights into the validity and potential perspectives of this urban living concept. Indeed, it can be a valuable tool for place-based policies aimed at correcting patterns of uneven spatial development, particularly in the design of redistributive interventions to improve access to essential services. As a second contribution, we perform a network theory exercise to model urban interactions within the city. In particular, we translate accessibility relationships between buildings and services into a weighted and directional network and apply a clustering algorithm to identify urban communities wined together by sharing the same services. Identifying service-based communities enables the analysis of phenomena such as urban segregation using endogenously generated communities rather than administrative boundaries. Finally, we propose a functional redundancy indicator to assess the resilience of the previously calculated \textit{10-minute accessibility index}. \\

\noindent The rest of the paper is organized as follows. Section \ref{sec: data} introduces the data used in the analysis. Section \ref{sec: 10min} illustrates the \textit{10-minute Index} and shows an application to the case of Florence. Section \ref{sec: network} presents the network theory application, while Section \ref{sec: redund} is dedicated to the redundancy issue. Finally, Section \ref{sec: concl} provides concluding remarks.


\section{Data}
\label{sec: data}

To construct the indicator, we need the map of residential buildings, as units of analysis, and the set of points of interest (POIs), representing the locations of essential services. \\

\noindent We retrieve the complete map of Florence's buildings from Open Street Maps (OSM) and exclude buildings that are certainly not intended for residential use. We refine the building sample using OSM's building classification tags, which indicate for each building its purpose. We then exclude educational and university facilities, hospitals, institutional buildings hosting public authorities, barracks, churches, convents, and prisons. Additionally, we remove buildings located in industrial areas, which are distinctly marked on the OSM map. To conclude, we drop buildings with an area below 28 $m^2$, the minimum size required by Italian habitability standards for a studio apartment intended for a single person\footnote{This is established by the Ministerial Decree of July 5, 1975, No. 75. \textit{Ministero dei Lavori Pubblici. Decreto Ministeriale 5 luglio 1975, n. 75, Norme per l’edilizia abitativa, Gazzetta Ufficiale della Repubblica Italiana, n. 235, 16 agosto 1975.}}. We end up with a sample of 37,950 residential buildings. \\

We select the points of interest by focusing on a set of essential services for daily lives of citizens. We consider 25 typologies of services which we group into six categories as follows:

\begin{itemize}
    \item \textbf{Schools}. This category includes the following types of services: nurseries, kindergartens, primary schools, and secondary schools. 
    \item \textbf{Healthcare}. This includes the following types of services: emergency rooms, medical clinics, and general practitioners' offices.
    \item \textbf{Primary services}. This includes the following types of services: tobacco shops, universal postal services, banks, and drugstores. 
    \item \textbf{Green Areas}: accessible public green areas. 
    \item \textbf{Leisure}. This includes four subcategories: \textit{associative activities} including the following types of services: Italian cultural and recreational clubs and places of worship; \textit{cultural activities} including the following types of services: theaters, cinemas, museums and libraries; \textit{sport activities}  including the following types of services: sport centers, school gyms and equipped sport areas; \textit{children activities}  including the following types of services: community play centers and equipped playgrounds.
    \item \textbf{Food retail}. This includes the following types of services: supermarkets and open-air markets.
\end{itemize}

While OSM is a commonly used source for collecting data points on urban amenities, we choose not to rely on it since OSM operates as a user-contributed platform that shows inconsistencies and gaps in data points coverage. As a result, the completeness and reliability of its information may vary significantly across different locations, making it less suitable for our purposes. We retrieve the geographical location of services, up to a total of 2,206 destinations, either in the form of geographic coordinates or postal addresses, from multiple official sources, such as institutional websites, offering complete lists of facilities for the city of Florence (see Table \ref{tab:POI_sources} for a detailed list of sources of geographic information by typology of service). The data collection is completed by georeferencing postal addresses with the Google Map API.\\

\begin{table}[H]
\centering
\caption{Sources of geographical data for services}
\label{tab:POI_sources}
\resizebox{\textwidth}{!}{%
\begin{tabular}{llc}
\hline \hline
& & \\
\textbf{Services} & \textbf{Data Sources} & \textbf{Geocoding} \\ 
& & \\[-1.5ex] \hline
& & \\
\begin{tabular}[c]{@{}l@{}} Schools, Museums, Green Areas, \\
Drugstores, Playgrounds  \end{tabular} & \url{https://opendata.comune.fi.it/}  & No\\
& & \\
\begin{tabular}[c]{@{}l@{}} Open-air markets, Libraries, \\ Sport activities, Community play centers \end{tabular}  & \url{https://www.comune.fi.it/}  & Yes\\
& & \\
Medical clinics & \begin{tabular}[c]{@{}l@{}} \url{https://www.misericordia.firenze.it/} \& \\ \url{https://www.uslcentro.toscana.it/} \end{tabular} & Yes \\
& & \\
\begin{tabular}[c]{@{}l@{}} Emergency rooms, \\ General practitioners' offices \end{tabular} & \url{https://www.uslcentro.toscana.it/} & Yes \\
& & \\
Tobacco Shops & \url{https://acciseonline8.adm.gov.it} & Yes \\
& & \\
Post Offices &  \url{https://www.poste.it/} & Yes \\
& & \\
Banks & \url{https://www.comuni-italiani.it/048/017/banche/} & Yes \\
& & \\
\begin{tabular}[c]{@{}l@{}} Supermarkets, Places of worship, \\ Cinemas, Theaters \end{tabular} & \url{https://www.openstreetmap.org/} & No\\
& & \\
ARCI Clubs\footnote{\textbf{definition of arci clubs}} & \url{https://www.arci.it/} & Yes \\ \hline
\end{tabular}%
}
\end{table}

\section{The \textit{10-minute Index}}
\label{sec: 10min}

The \textit{10-minute Index} measures how easily people in a city can reach essential services within a 10-minute walk. With respect to the traditional concept of \textit{15-minute city}, we introduce an adjustment to the travel time threshold by bringing it down to 10 minutes. This change is made for a major reason. From a methodological perspective, our approach introduces an unprecedented level of precision in defining starting points of pedestrian routes. Unlike other methodologies that often rely on administrative area centroids, we directly refer to the buildings where people live. This permits minimizing errors in determining the actual starting point of a journey and enables us to adopt a stricter travel-time threshold. According to our definition, \textit{accessibility to a service for a citizen exists when the service is within a 10-minute walking distance from their home.} \\

We develop the \textit{10-minute Index} using a rigorous spatial analysis process enabled by Geographic Information Systems (GIS) and open-source geospatial data. Our methodology consists of the following steps: \\

\noindent \textit{a) Find buildings' centroids.} After identifying the residential buildings, as detailed in Section \ref{sec: data}, we mark the centroid of each building, which serves as the starting point for accessibility calculations. \\

\begin{figure}[H]
    \caption{Delineating building-centered isochrones}
    \label{fig: 10_min_meth}
    \centering
    \begin{subfigure}[t]{0.48\textwidth}
        \centering
        \includegraphics[width=\linewidth]{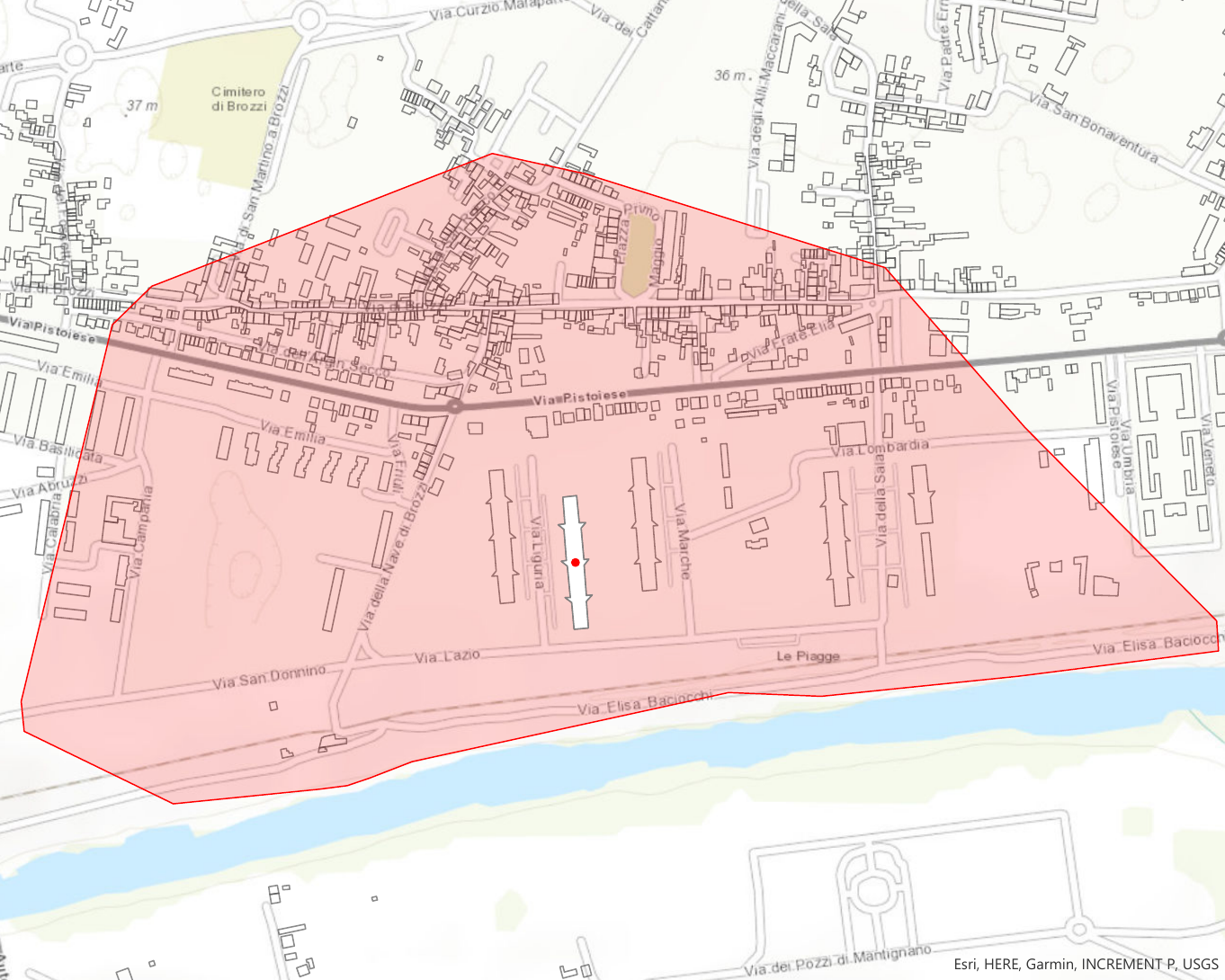} 
                    \caption{Generating a building's isochrone} \label{fig: panela}
    \end{subfigure}
    \begin{subfigure}[t]{0.48\textwidth}
        \centering
        \includegraphics[width=\linewidth]{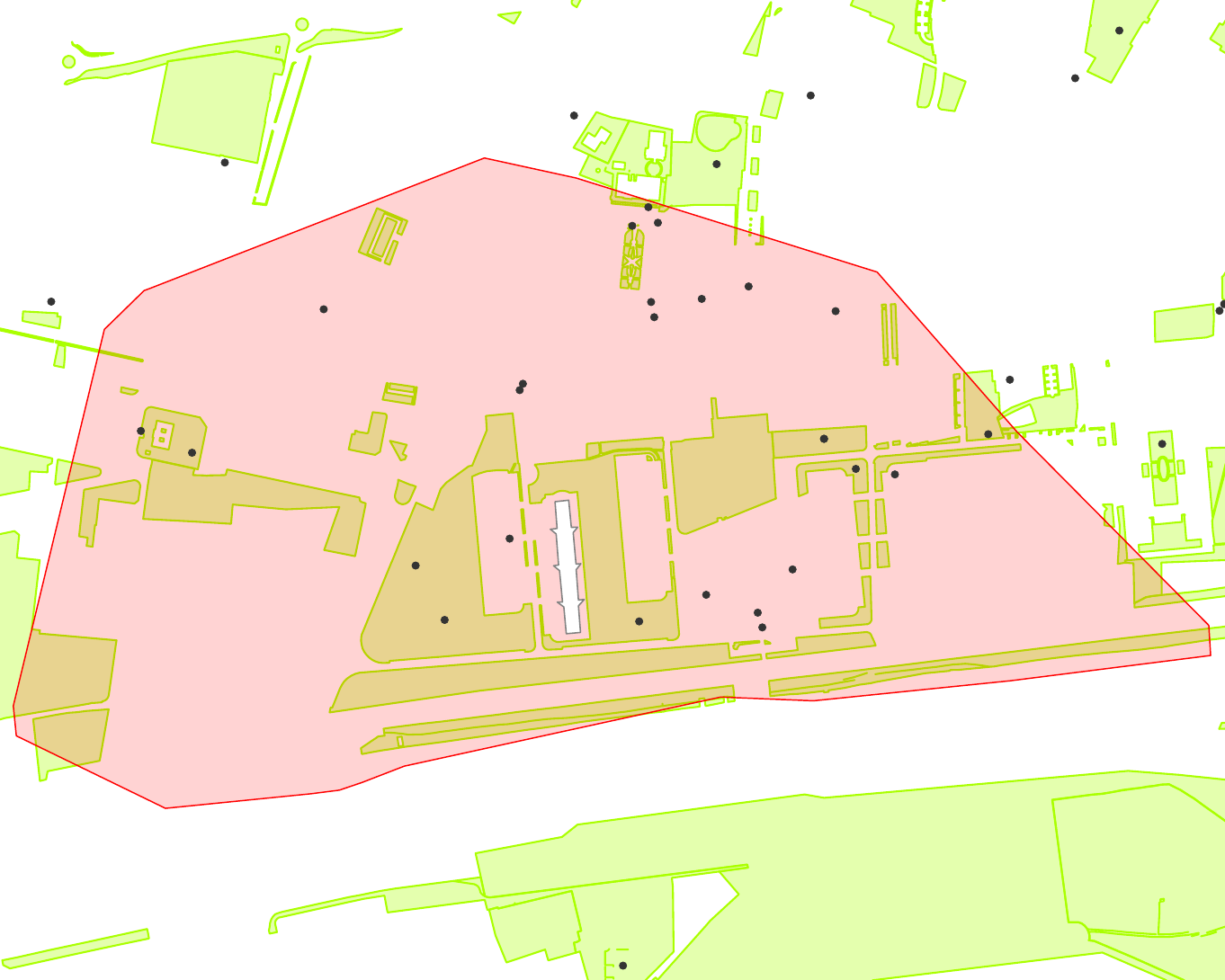} 
                    \caption{Overlapping the isochrone with POIs} \label{fig: panelb}
    \end{subfigure}
    \begin{tablenotes}
      \scriptsize 
      \item Notes: This figure illustrates the process of identifying an isochrone and the intersecting POIs for a real building on the map of Florence. The same process is applied to all residential buildings in our sample. The building in question is located in the Quaracchi suburb, on the western outskirts of Florence, and is administratively part of the Rifredi district (Quartiere 5). Panel a) shows the selected building, with its centroid (the red point) and isochrone (the red-highlighted area), generated using ORS. Panel b) shows the overlap between the isochrone and the POIs (represented by blue points): once the isochrone is identified, essential services, including green areas (depicted as green polygons), are projected onto it, and only those that fall within the isochrone are considered accessible from the building. 
\end{tablenotes}
\end{figure}

\noindent \textit{b) Generate buildings' isochrones.} We generate the 10-minute walk isochrone for each building's centroid using the OpenRouteService (ORS) API. The latter is a routing tool that relies on street network data of OpenStreetMap\footnote{The OSM platform collects and provides geospatial information contributed by a global community and has now reached a high level of coverage and reliability. Note that \cite{barrington2017world} find an approximately 80\% completeness of OSM road data and that completeness has a U-shaped relationship with population density, meaning that sparsely populated areas and dense cities are the best mapped.}. An isochrone is the polygon delineated by all points accessible within a 10-minute walk from a given building according to ORS, which assumes a walking speed of 5 km/h. Figure \ref{fig: panela} illustrates an example of what an isochrone\footnote{Note that we exclude isochrones when the corresponding building centroid falls outside the perimeter. This happens when the building is located off a walkable route (e.g., in isolated rural areas). In these cases, OSM repositions the starting point to the nearest walkable route, which may, in some instances, fall outside the isochrone.} looks like.\\

\noindent \textit{c) Overlap isochrones and points of interest.} For each isochrone, we identify the intersecting POIs, which represent the essential services reachable within a 10-minute walk from a given building. This step is illustrated in the example in Figure \ref{fig: panelb}, where the blue points represent essential services and the red area is the isochrone of a randomly selected building on the map. The blue points within the isochrone are considered accessible from the building around which the isochrone is generated.\\

\noindent \textit{d) Calculate type-specific accessibility.} Accessibility is evaluated separately for each type of service. If at least one POI of a specific type intersects the isochrone, the building is considered to have access to that specific service. Green areas require a separate discussion. In this case, accessibility is evaluated on a scale from 0 to 1, depending on the portion of public green space intersected by the isochrone. \\

\noindent \textit{e) Aggregate into category-specific score.} As seen in the previous section, service types are organized into categories, and in some cases, first into subcategories before being grouped into categories. For clarity, we now outline how the score for each of the six categories is calculated for each building:\\
\begin{itemize}
    \item In the case of \textit{School}, \textit{Healthcare}, \textit{Primary Services}, and \textit{Food Retail}, the score consists of the ratio of accessible service types within the category to the total number of service types in the category. If no service type in the category is accessible, the score for that category is 0. If at least one service of every type is accessible, the score is 1.
    \item In the case of \textit{Leisure}, a sub-score is calculated for each subcategory using the same method as above. The overall \textit{Leisure} score is the average of these sub-scores. Thus, sub-scores range from 0 (no accessible service types) to 1 (all service types accessible).
    \item In the case of \textit{Green Areas}, we look at the intersection between each isochrone and green spaces. If the intersection exceeds a predefined threshold, the score is set to 1, the maximum value. If it falls below the threshold, the score is determined by dividing the intersection area by the threshold, yielding a value between 0 and 1. To set this threshold, we reviewed the recommendations of the World Health Organization (WHO) and guidelines adopted by several major European cities. According to the WHO publication \textit{Urban green spaces: a brief for action} \citep{europe2017urban}, it is recommended that urban residents have access to at least 0.5–1 hectare of public green space within 300 meters of their home which corresponds to a 4-minute walk. Many cities have adopted and adapted the recommendations\footnote{In the 2020 document \textit{Provision of Residential Nearby Public Green Spaces}, the Senate Department for Urban Development and Housing outlines a guideline for the city of Berlin: residents should have access to green spaces within 500 meters to spaces of at least 0.5 hectares, and 1–1.5 kilometers to larger green spaces of at least 10 hectares. The city of Barcelona follows a policy, formalized in the \textit{Green Infrastructure and Biodiversity Plan}, that aims to ensure all its residents have access to green space within 10 minutes of walking.}. These guidelines are based on linear distances, but we adapt them to our context, where geographic areas are examined. Specifically, we follow the criteria provided by Natural England, the UK government agency responsible for protecting and conserving the natural environment\footnote{The \textit{Accessible Natural Greenspace Standard} (ANGSt), established by Natural England, recommends that people should have access to at least 2 hectares of green space within 300 meters of their home (full document available \href{https://designatedsites.naturalengland.org.uk/GreenInfrastructure/GIStandards.aspx}{here}). These guidelines are then incorporated into the urban planning of cities across England.}, calculating the equivalent for areas defined by our isochrones. We eventually set the threshold at 8 hectares, above which the maximum accessibility score of 1 is assigned.
\end{itemize}

\noindent \textit{f) Aggregate into the 10-minute Index.} Each building receives an overall score, i.e. the 10-minute Index, given by the average of the six type-specific scores. This final indicator reflects the degree of accessibility across all considered categories. \\

\noindent Our methodology is specifically designed to provide an independent evaluation of each building. Each building is assigned a value based on its own features, which does not depend on the performance registered in other areas of the city, ensuring an absolute measure of accessibility. This approach is preferred as it makes results easier to understand and interpret without accounting for the varying conditions or performance of other buildings. Additionally, it allows planners and decision-makers to identify specific areas where buildings may need improvement, regardless of how other buildings perform. 

\subsection{An application to the city of Florence}

In this subsection, we introduce the \textit{10-minute Index} for the city of Florence. To provide context, we first highlight key aspects of Florence’s demographics and urban morphology. Furthermore, examining specific characteristics of the city helps support our proposed new lower threshold of 10 minutes. \\

The city of Florence exhibits features that classify it as a \textit{compact} city. The term \textit{compact} refers to an urban form characterized by high-density and proximate development, where land is efficiently utilized, and urban areas are well-integrated \citep{oecd2012compact}. These attributes are beneficial because higher density reduces dependence on automobiles, while mixed land use ensures diversity in service supply, allowing most residents to access essential services on foot or via soft transport options. \\

Density is one of the key factor in defining a city as \textit{compact}. As of 2011, the municipality of Florence covers an area of 102,31 Km$^2$. With a population density of about 3,535 people per square kilometer, Florence has a moderately high density compared to other major Italian cities. It ranks 50th among all Italian municipalities\footnote{For perspective, its density is lower than cities like Naples (7,745 people/km²) and Milan (7,537 people/km²) but higher than Rome (2,134 people/km²) and Genoa (2,357 people/km²).}. Moreover, the metropolitan city of Florence, which is part of the 21 Italian local systems identified as main urban areas by \cite{istat2017urban}\footnote{These 21 main urban areas are identified according to the following criteria: being in a metropolitan city, a local system population of over 500,000, or a main municipality population over 200,000. Based on these, the selected local systems are: Turin, Busto Arsizio, Como, Milan, Bergamo, Verona, Venice, Padua, Trieste, Genoa, Bologna, Florence, Rome, Naples, Bari, Taranto, Reggio Calabria, Palermo, Messina, Catania, and Cagliari.} is covered for a 11.8\% by inhabited areas, which is a significantly higher share with respect to the national average of 6.7\% \citep{istat2017urban}. 

\begin{figure}[H]
    \caption{Sub-city areas of Florence}
    \label{fig: subcity}
    \centering
        \includegraphics[width=.9\linewidth]{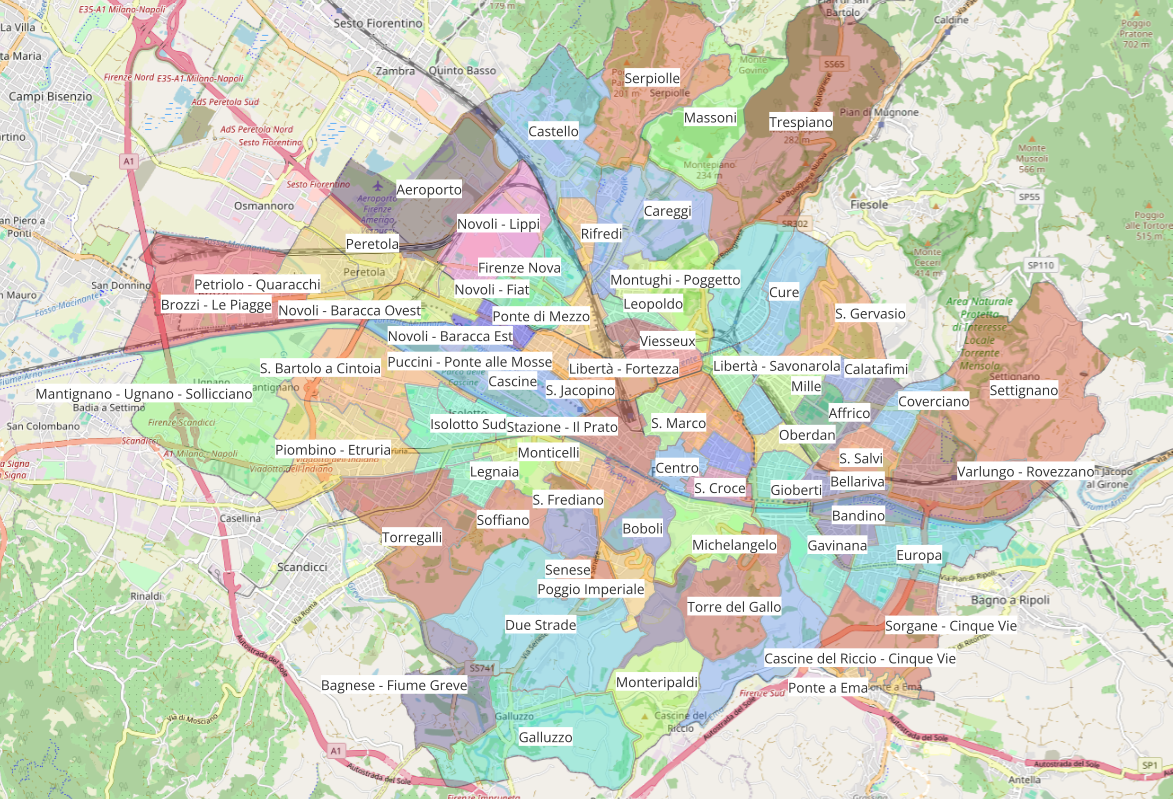} 
    \begin{tablenotes}
      \scriptsize 
      \item Notes: The picture above illustrates the map of Florence from OSM divided into the 72 elementary areas defined by ISTAT, each labeled with its corresponding designation. Elementary areas are sub-municipal units at an intermediate level between census tracts and administrative subdivisions.
\end{tablenotes}
\label{fig: index_map}
\end{figure}

One more aspect is how this density is distributed within the municipality. If urbanization is continuous, with high-density areas spread evenly across the city, then the city might be considered \textit{compact}. Conversely, if there are sparsely inhabited areas interrupting this continuity, it challenges the city's \textit{compactness}. To do this, we analyze the population density of the 72 elementary areas as defined by ISTAT and reported in Figure \ref{fig: subcity}. At the core of the city lies the elementary area \textit{Centro}, as shown in Figure \ref{fig: subcity}, which is surrounded, in a ring-like pattern, by the elementary areas making up Florence's historic center. The elementary areas with the highest population density in Florence are Leopoldo (22,593 inhabitants per $km^2$), followed by Puccini - Ponte alle Mosse (19,878), Novoli - Baracca Est (18,485), Viesseux (17,852), and Rifredi (16,638). Other highly populated areas include Calatafimi (15,287), S. Spirito (15,162), Legnaia (14,979), and Oberdan (14,247). While some of these areas are centrally located, others are situated in more peripheral parts of the city (see Novoli - Baracca Est, Rifredi and Calatafimi in Figure \ref{fig: subcity}). At the lower end of the ranking, we find the least densely populated elementary areas in Florence. Cascine del Riccio - Cinque Vie (507 inhabitants per square kilometer) is the most densely populated among them, followed by Monteripaldi (469), Torre del Gallo (420), Settignano (413), and Trespiano (289). Further down the list are Serpiolle (286), Arcetri (253), Bagnese - Fiume Greve (226), Massoni (109), and finally Aeroporto (69), which has the lowest population density. 
Notably, the least populated areas are concentrated along the city's outer edges, rather than being scattered among densely populated zones, and as a result, there are no major gaps in urban population density. This characteristic further support Florence's classification as a \textit{compact} city. \\ 

These observations are significant as they indicate that Florence is likely to achieve good service accessibility. Now, let us analyze how the \textit{10-minute Index} is distributed across the city. Figure \ref{fig: index_map} presents the values obtained for Florence.

\begin{figure}[H]
    \caption{The \textit{10-minute index} in Florence}
    \centering
        \includegraphics[width=.9\linewidth]{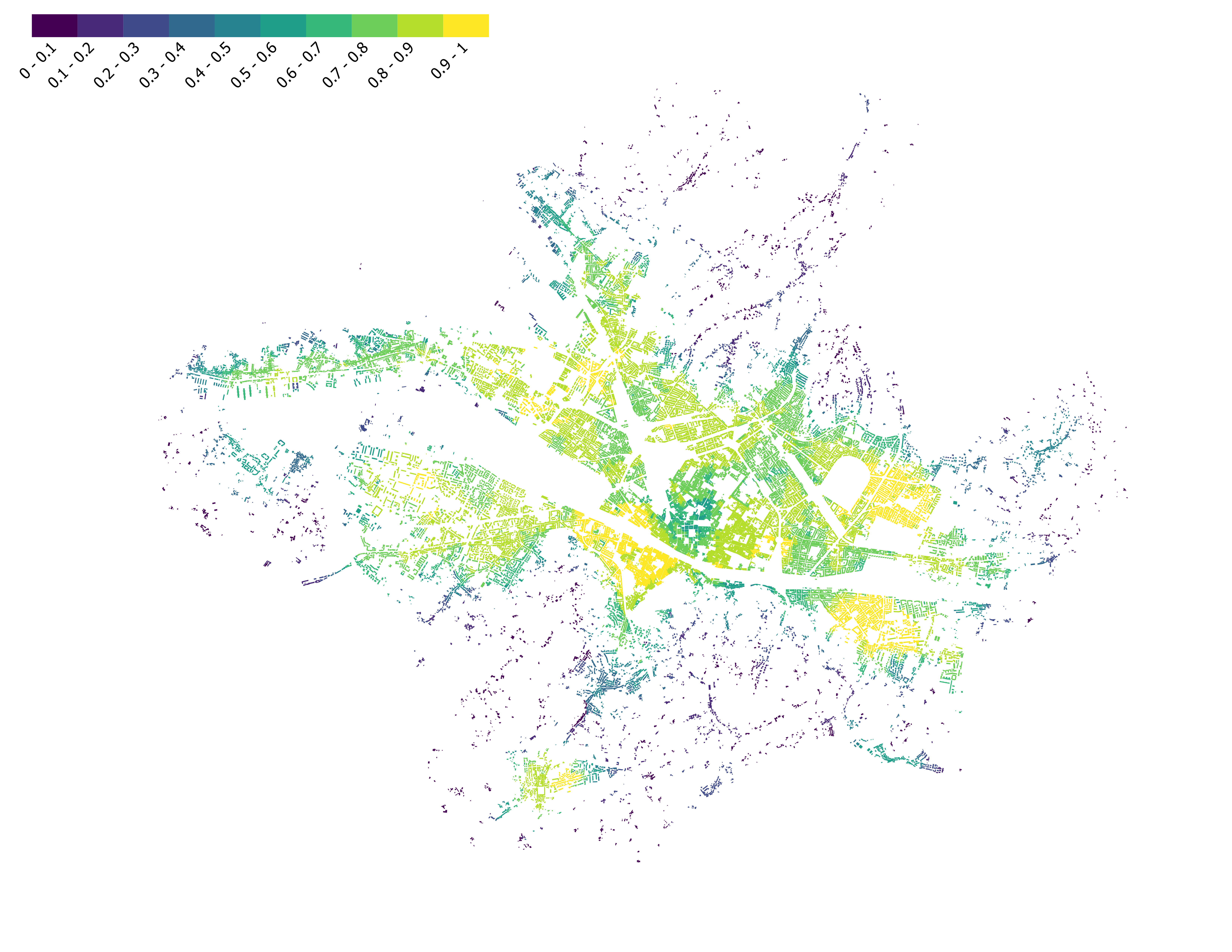} 
    \begin{tablenotes}
      \scriptsize 
      \item Notes: This map displays the buildings in Florence, where each building is color-coded according to its \textit{10-minute index} value. The colors range from blue, indicating the lowest values, to yellow, representing the highest values.
\end{tablenotes}
\label{fig: index_map}
\end{figure}

Three larger zones stand out where the index remains consistently high. The first is centrally located, covering the areas of Pignone, S. Frediano, and S. Spirito. The other two are outside the city center: one to the southwest, encompassing Bandino/Bellariva, and another to the northwest, spanning Affrico/Coverciano. Additionally, there are smaller high-performing areas, as indicated by the small yellow patches, such as those in the south near Galluzzo and in the northeast around Novoli. The areas with the lowest performance are the least populated ones, where housing is more scattered due to the natural layout of the territory. However, while low residential density is clearly associated with very low index values, it is important to note that a higher population density does not necessarily correspond to the highest possible index values. \\

\begin{figure}[H]
    \centering
    \caption{Category-specific scores}
    \label{fig: subind}
    \label{fig:6images}

    \begin{subfigure}[b]{0.3\textwidth}
        \centering
        \includegraphics[width=\textwidth]{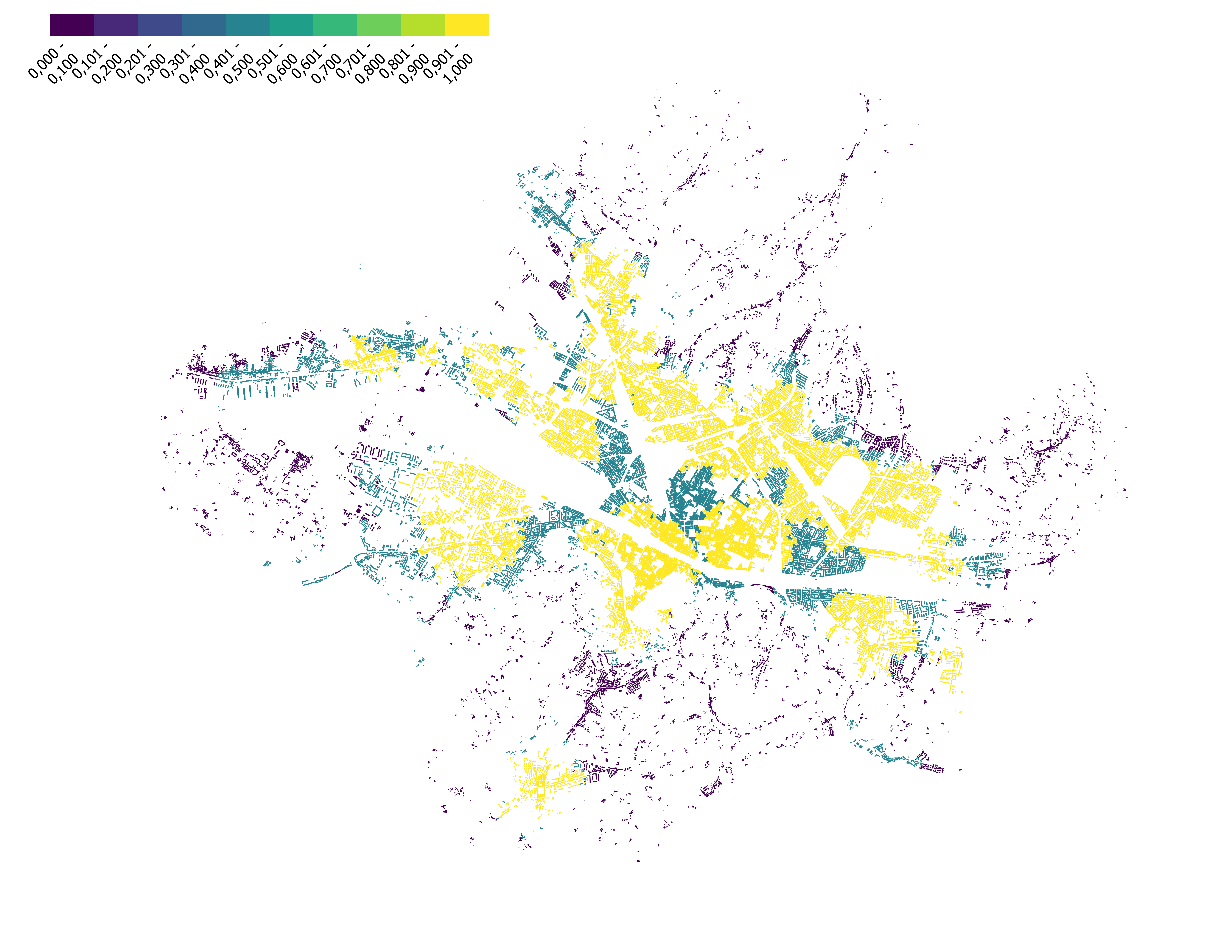}
        \caption{Food Retail}
        \label{fig:image1}
    \end{subfigure}
    \hfill
    \begin{subfigure}[b]{0.3\textwidth}
        \centering
        \includegraphics[width=\textwidth]{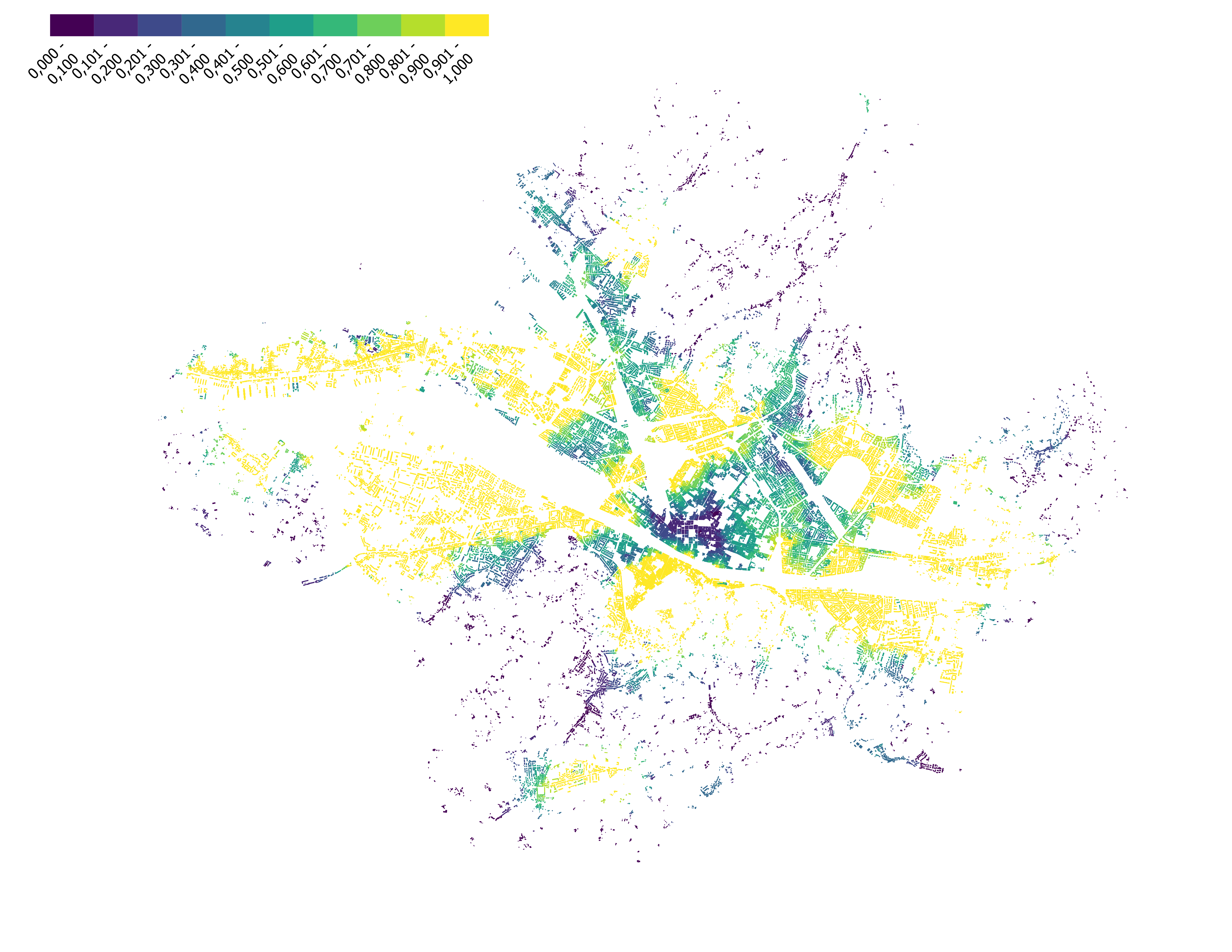}
        \caption{Green Areas}
        \label{fig:image2}
    \end{subfigure}
    \hfill
    \begin{subfigure}[b]{0.3\textwidth}
        \centering
        \includegraphics[width=\textwidth]{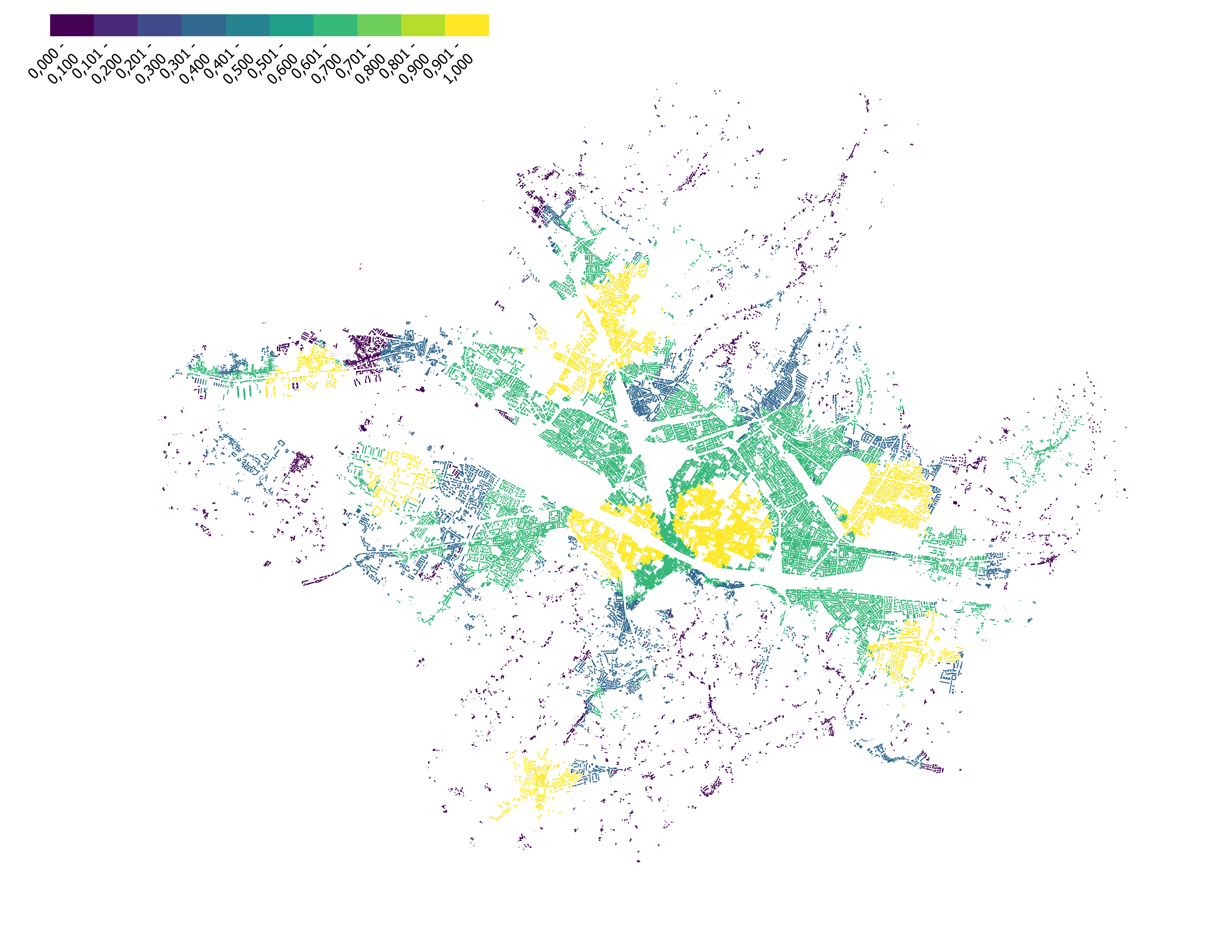}
        \caption{Healthcare}
        \label{fig:image3}
    \end{subfigure}

    \vskip\baselineskip 

    \begin{subfigure}[b]{0.3\textwidth}
        \centering
        \includegraphics[width=\textwidth]{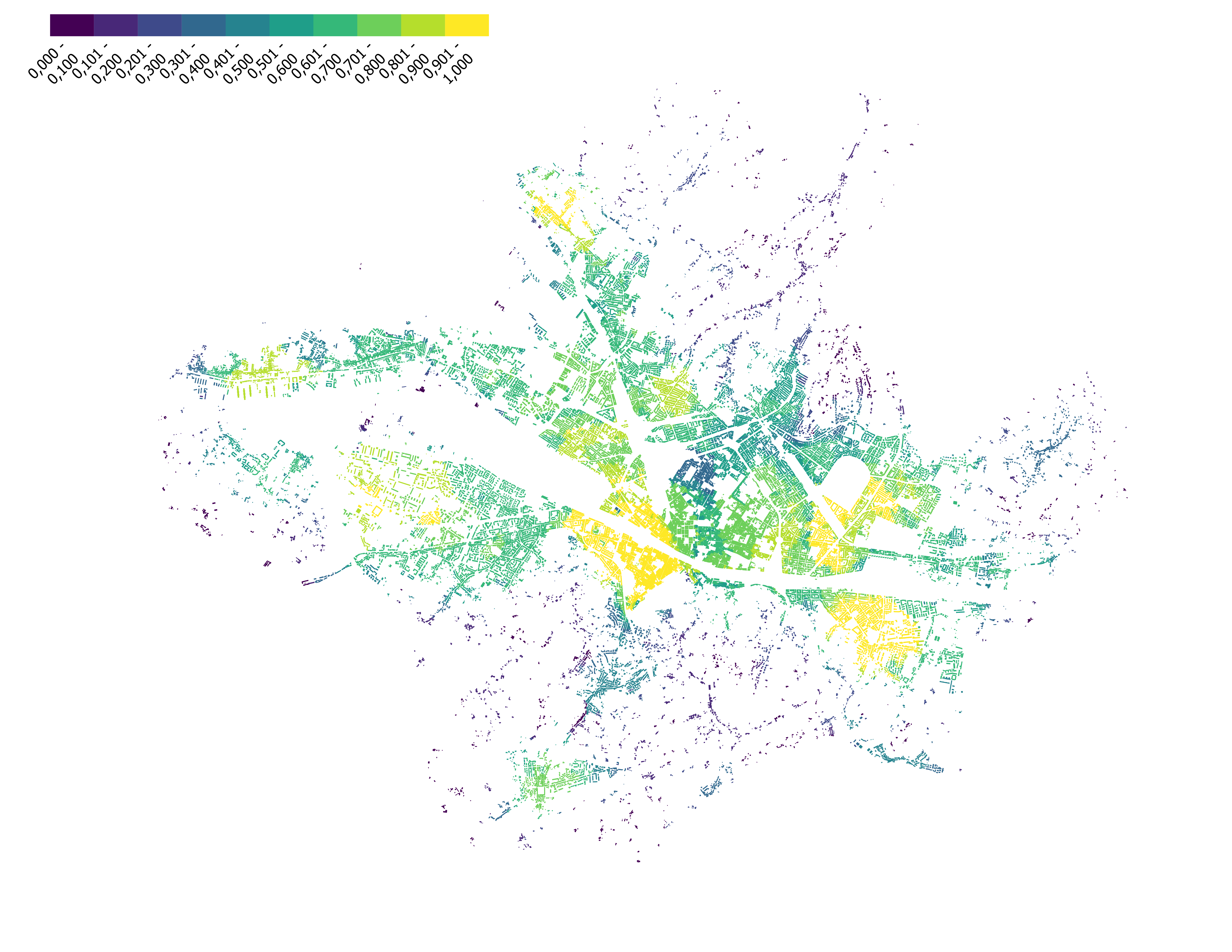}
        \caption{Leisure}
        \label{fig:image4}
    \end{subfigure}
    \hfill
    \begin{subfigure}[b]{0.3\textwidth}
        \centering
        \includegraphics[width=\textwidth]{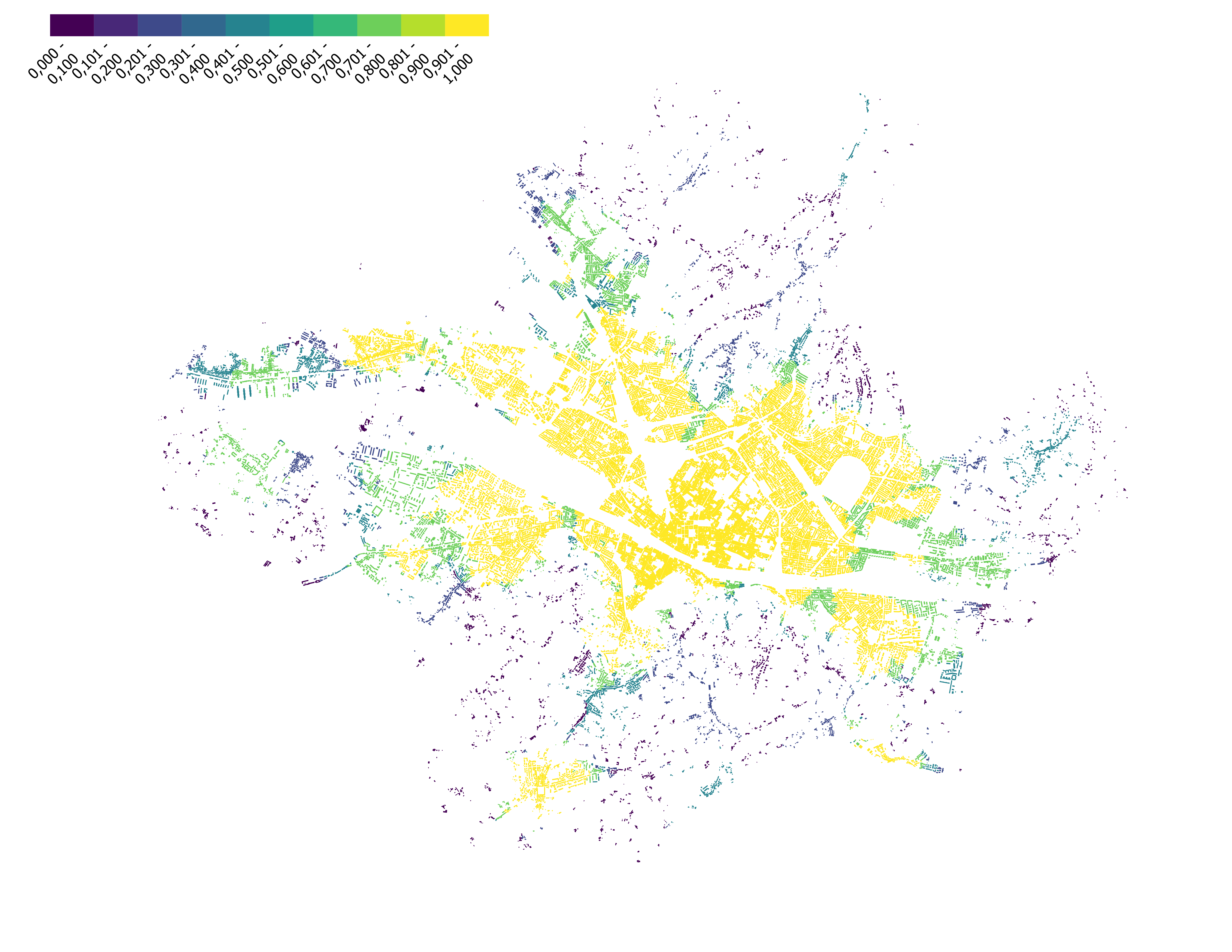}
        \caption{Primary Services}
        \label{fig:image5}
    \end{subfigure}
    \hfill
    \begin{subfigure}[b]{0.3\textwidth}
        \centering
        \includegraphics[width=\textwidth]{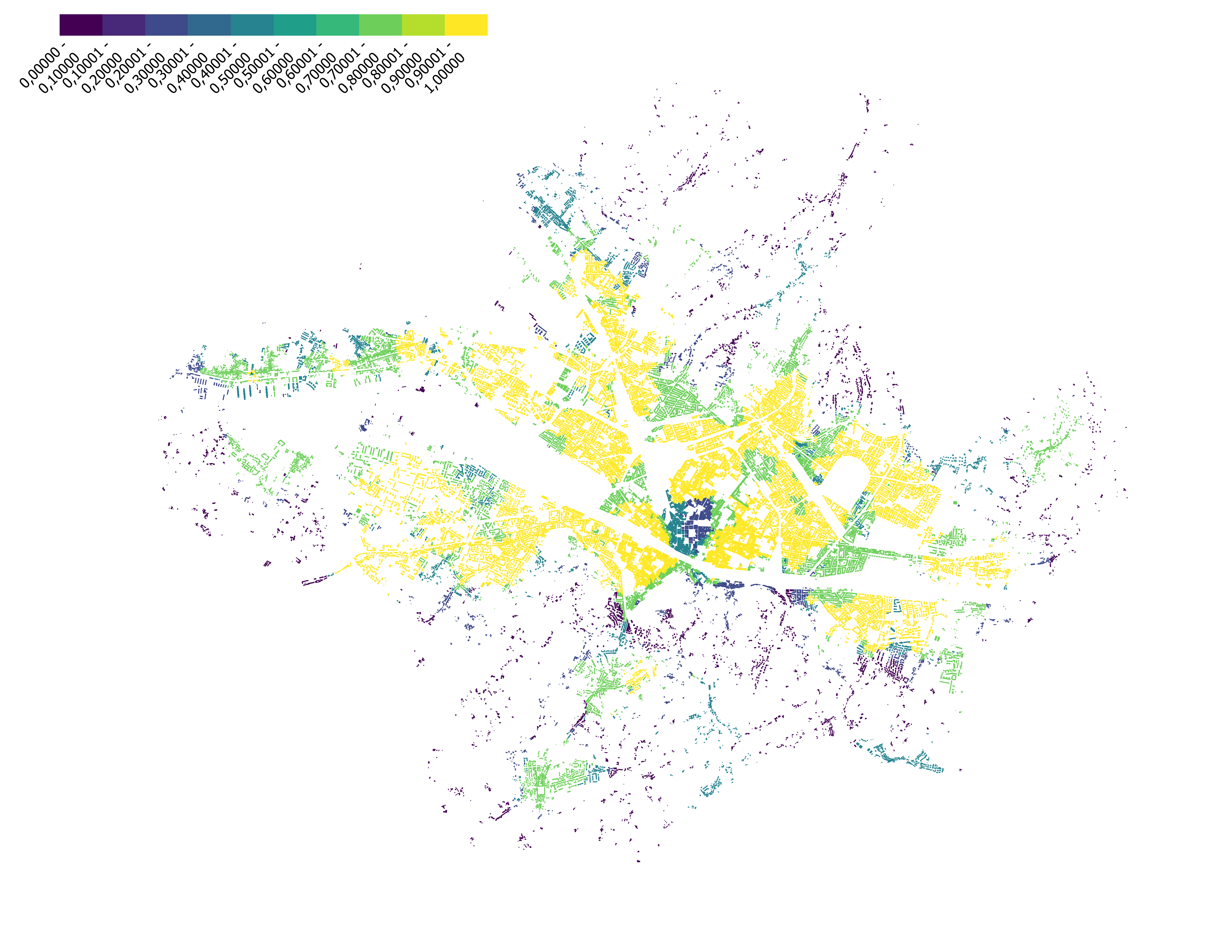}
        \caption{Schools}
        \label{fig:image6}
    \end{subfigure}
     \begin{tablenotes}
      \scriptsize 
      \item Notes: Each map displays the buildings in Florence, where each building is color-coded according to its category-specific score. The colors range from blue, indicating the lowest values, to yellow, representing the highest values.
\end{tablenotes}   
\end{figure}

Figure \ref{fig: subind} shows the score assigned to residential buildings by service category. We observe that half of the observation units have access to all categories of primary services, food retail, and schools. Half of the buildings have access to at least two types within the healthcare category. 25\% of the buildings have access to an optimal amount of green space. Only 5\% have access to all types of leisure services. \\

\section{Community of services}
\label{sec: network}

In this section, we propose a methodology to identify spatial clusters of services in urban areas. 

\begin{figure}[H]
    \caption{Building the adjacency matrix}
    \centering
        \includegraphics[width=1\linewidth]{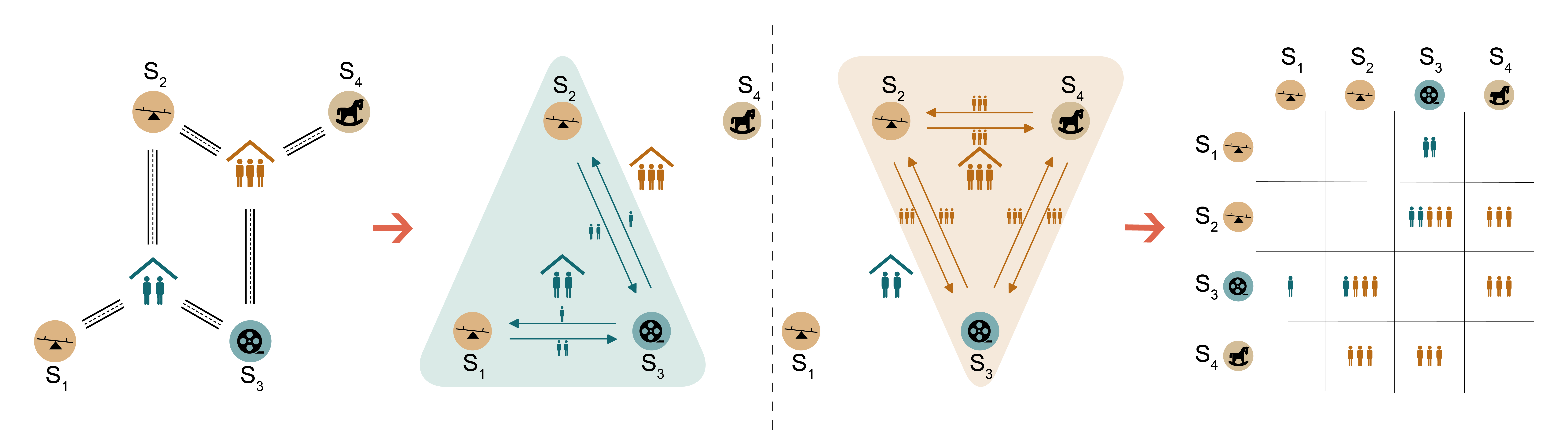} 
    \begin{tablenotes}
      \scriptsize 
      \item Notes: The picture above consists of three sequential panels, each depicting a step in the process of constructing the matrix used for applying the community detection algorithm. Consider a fictional city with two buildings: one housing two individuals (blue house) and the other three (red house). The city includes a total of four services of three typologies: two playgrounds ($s_1$ and $s_2$), a cinema ($s_3$), and a kindergarten ($s_4$), as shown in the first panel. In the second panel, the blue building has access to $s_1$, $s_2$, and $s_3$, as these services fall within its isochrone (the blue-shaded area). Meanwhile, the red building has access to $s_2$, $s_3$, and $s_4$, as they are covered by its isochrone (the red-shaded area). We assume that residents move from one service to another within their isochrone, using the services according to the rules defined by equation \ref{eq_sistema_pesi}. To illustrate this, we refer to the blue building in the second panel as an example. Its residents can reach two playgrounds and a cinema within a 10-minute walk. Based on equation \ref{eq_sistema_pesi}, we make the following assumptions: i) residents do not use the same type of service more than once, so they will not visit $s_1$ after visiting $s_2$ and viceversa. Instead, there will be movement between services of different types, as shown by the double arrows between $s_1$ and $s_3$, and $s_2$ and $s_3$; ii) when there are multiple services of the same type, residents will evenly distribute between them. Therefore, the movement from service $s_3$ to $s_1$ and $s_2$, which are of the same type, involve only one of the two residents of the blue building; iii) when there is a unique service, like $s_3$, residents will share it. The third panel displays the $4\times4$ service matrix, where each entry represents the total population that utilizes two services together, physically traveling from one to the other.

\end{tablenotes}
\label{fig: matrix_procedure}
\end{figure}

Urban interactions can be effectively modeled using network representations, here, we aim to explore the relationships between different services in terms of 10-minute walk accessibility. Specifically, we focus on identifying services that are shared by the same group of beneficiaries. By beneficiaries, we mean individuals residing in buildings whose isochrones encompass these services. To achieve this, we model the services as the nodes of a weighted and directed network. The directional and weighted edges represent the population that, as potential beneficiaries of the originating service, are also potential beneficiaries of the destination service. Additionally, we account for the category of each service, assuming that beneficiaries of a service do not require access to other services of the same category. Formally, we have that the set of nodes in the network is represented by the set of services $S=\left\{ s_{1},\dots ,s_{N} \right\}$. We must therefore construct the adjacency matrix of the weighted and directed network, which will be an
$N\times N$ matrix where the value $w_{ij}$, in entry $(i,j)$, represents the weight of the edge starting from service $s_i$ and ending in service $s_j$. Each service $s_i \in S$ is associated with a specific typology. The set of typologies is given by $T=\left\{ t_1,\dots,t_m \right\}$. Let $\Gamma$ be a function that maps the services to their corresponding typology:
\begin{equation*}
    \Gamma: S \rightarrow T  
\end{equation*}
The weight of each edge $w_{ij}$ is determined by the sum of contributions from buildings that provide access to both services $s_i$ and $sj$. Let $b^{\alpha}$, with $\alpha \in [1,B]$, be a building with population $P^{\alpha}$ and $S^{\alpha}$ be the subset of services that are contained in the isochrone of building $b^{\alpha}$. Then, the contribution of building $b^{\alpha}$ to the weight $w_{ij}$ is:
\begin{equation}\label{eq_sistema_pesi}
w_{ij}^{\alpha} = \left\{
\begin{array}{l cl}
    0 && \text{if } \Gamma(S_i)=\Gamma(S_j) \\
    0 && \text{if } S_i \notin S^{\alpha} \text{ or } S_j \notin S^{\alpha} \\
    P^{\alpha}/n^{\alpha}(\Gamma(s_j)) & &\text{otherwise}
\end{array}
\right.
\end{equation}
Where $n^{\alpha}(\Gamma(s_j))$ is the number of services in $S^{\alpha}$ with the same typology of the service $s_j$:
\begin{equation*}
    n^{\alpha}(\Gamma(s_j)) = |\{ s\in S^{\alpha} : \Gamma(s)=\Gamma(s_j) \}|
\end{equation*}
We then obtain the weight of each edge as:
\begin{equation*}
    w_{ij} = \sum_{i=1}^{B} w_{ij}^{\alpha}
\end{equation*}
The adjacency matrix construction process is shown in Figure \ref{fig: matrix_procedure}.

\noindent As already seen in geographic analysis contributions, we employ a community detection algorithm to uncover urban clusters of services. For this purpose, we chose InfoMap \citep{rosvall2008maps}, a method that is well suited for weighted and directed networks, it has been shown to outperform other algorithms according to \citet{lancichinetti2009community} and offers a conceptual approach particularly suited to our context. InfoMap stands out due to its use of random walks as the basis for identifying communities, which provides an intuitive framework for analyzing networks of urban services where spatial and functional relationships are essential. In what follows, we provide a concise explanation of the algorithm’s functioning and explain why we prefer it over alternative community detection techniques. InfoMap's distinctive feature lies in its framing of the community detection problem as an issue of data compression. Specifically, it borrows principles from information theory to model the network as a series of movements by a random walker. The algorithm identifies communities by seeking partitions that minimize the length of the encoded description of these movements. In other words, it discovers groupings of nodes that allow for the most compact representation of a random walk, thereby revealing structures where transitions between nodes are concentrated within communities and less frequent across them. This focus on compressing random walks aligns well with our goal of identifying service clusters that reflect localized interactions and shared accessibility. 

In our analysis, we offer an interpretation of the algorithm's functioning that makes it particularly well-suited to our case. Rather than imagining an agent moving randomly across the network, we interpret the transition from one node to another as an individual moving from one service to the next. At the second service, this individual encounters another person who subsequently moves on to a different service, creating a chain of interactions. A cluster, as defined by the algorithm, is a group of nodes where a random walk, once it enters, tends to remain for an extended period. Under our interpretation, a cluster represents a set of services whose usage is shared by individuals who predominantly interact within the cluster.  \\

The next step in cluster analysis is moving from the service level to the building level, requiring the assignment of buildings to service communities. If a building's isochrone contains services from only one community, the assignment is easily done. However, if the isochrone includes services from multiple communities, the assignment is determined by evaluating the building's contribution to the internal links of each involved community. For every service community within the isochrone that includes at least two services, a score is calculated. Let $S^{\alpha,k}$ be the set of services contained in the isochrone of building $b^{\alpha}$ and belonging to community $k\in [1,K]$. Let $S^{\alpha,k}_{c_h}$, with $h\in[1,m]$ be the subset of $S^{\alpha,k}$ made of services with category $c_h$. The score of community $k$ for building $b^{\alpha}$ is:
\begin{equation}\label{eq_score_communities}
    \textit{score}^{\alpha,k} = \sum_{h=1}^{m} |S^{\alpha,k}|-|S^{\alpha,k}_{c_h}|
\end{equation}  
Finally building $b^{\alpha}$ is associated to the community with maximal score\footnote{Out of 37950 buildings, 55 are contested, i.e., two or more communities achieve the same highest score, and are thus unassigned; 1308 lack access to at least two different types of services and cannot be assigned to any community.}.  \\

\begin{figure}[H]
    \caption{Allocating border buildings to local communities}
    \label{fig: assign_to_comm}
    \centering
        \begin{subfigure}[t]{0.48\textwidth}
        \centering
        \includegraphics[width=\linewidth]{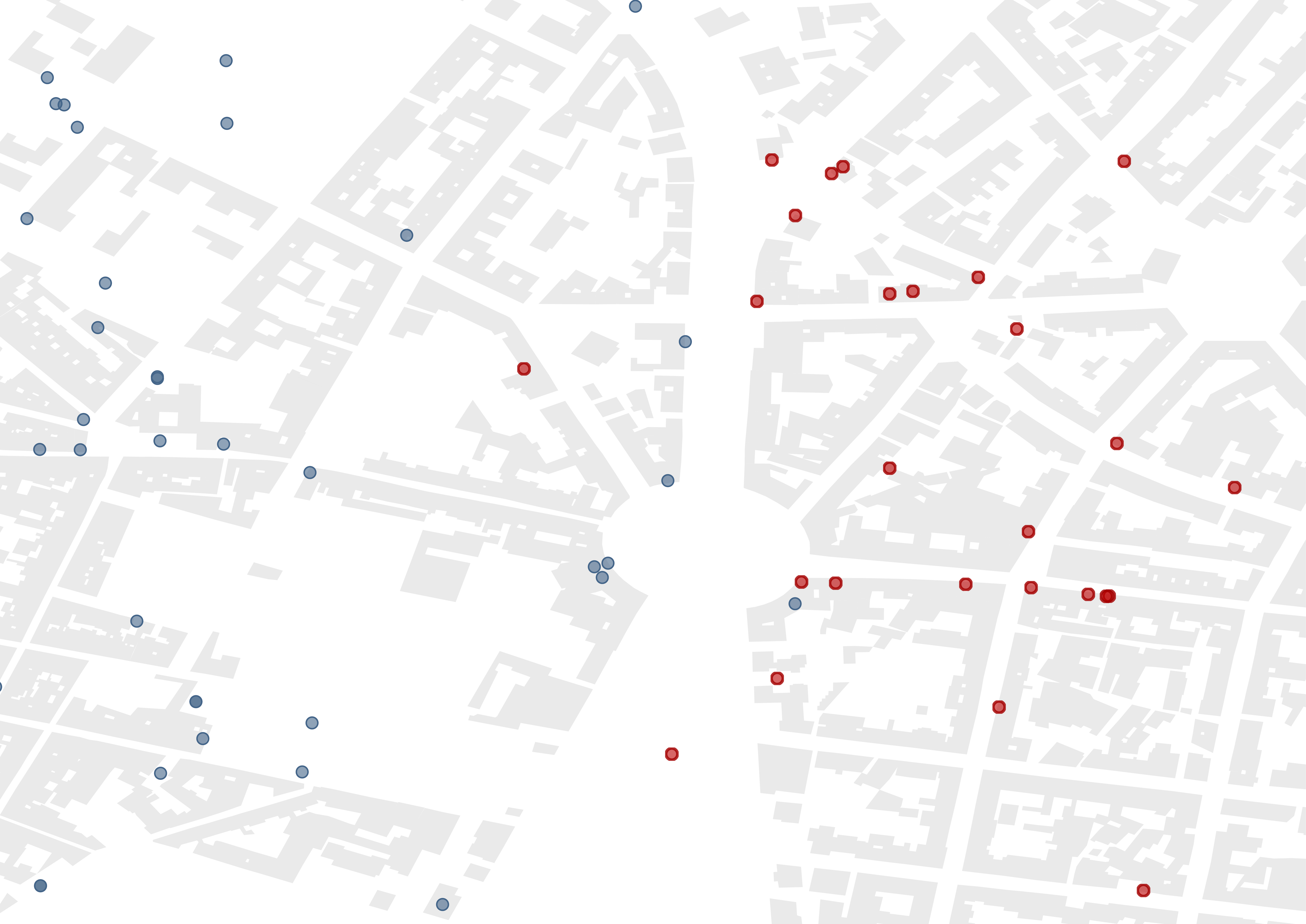} 
        \caption{} \label{fig: assign_to_comm2}
    \end{subfigure}
    \begin{subfigure}[t]{0.48\textwidth}
        \centering
        \includegraphics[width=\linewidth]{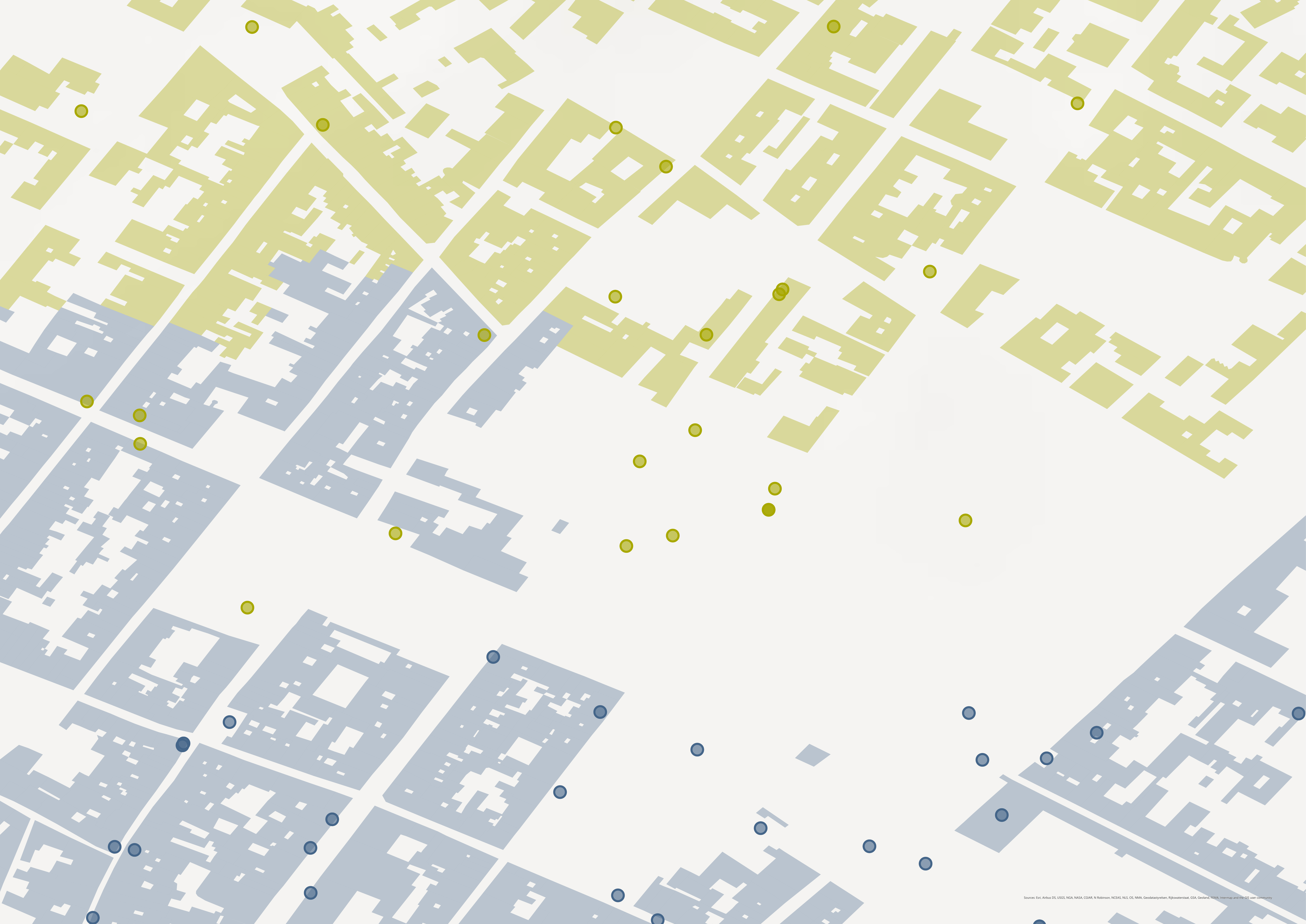} 
        \caption{} \label{fig: assign_to_comm1}
    \end{subfigure}
    \begin{tablenotes}
      \scriptsize 
      \item Notes: The figure above illustrates how buildings are assigned to communities in cases where they are located along the boundaries between two or more service communities. In panel (a), we zoom in on a section of the Florence map, focusing on buildings situated exactly at the border between two service communities, represented by red and blue points. In such cases, it is not immediately clear which community a building should belong to, as border buildings may have access to services from multiple neighboring communities. Panel (b) demonstrates how building assignments are determined after applying Equation \ref{eq_score_communities}. The allocation favors communities with greater service density and heterogeneity. As a result, we observe cases like the one in panel (b), where buildings assigned to one community (the blue one) are surrounded by services from an adjacent community (the green one).

\end{tablenotes}
\end{figure}

By applying the procedures previously explained for creating service communities and subsequently building communities, it is possible for some services within one community to be surrounded by buildings belonging to another community. As shown in Figure \ref{fig: assign_to_comm1}, this can occur in border areas between two communities. A preliminary explanation for this phenomenon lies in the fact that a service located inside a building has the same weight, in the procedures used to assign the building to a community, as a service located 10 minutes away on foot. Using concentric isochrones, i.e., calculating multiple isochrones for each building based on varying travel distances, would allow services to be weighted differently based on travel time, not just 10-minute accessibility. However, such a modification would not completely eliminate this outcome, which is partly desirable. Buildings are assigned to a community when the services within the isochrones of that community are relatively denser and more diverse than those of other communities, as indicated by Equation (\ref{eq_score_communities}). As a result, denser communities with greater service variability tend to attract buildings located within the convex hull of service communities from other areas. Another important point to note is that the convex hulls of neighboring communities may overlap, leading to outcomes not achievable through purely spatial clustering strategies (see Figure \ref{fig: assign_to_comm2}). During the community generation phase, in border areas, nearby services may be assigned to different communities based on the number of same-type services present in each community. If a service near the border is overrepresented in one community, it will form stronger links with services from the other community, as derived from the definition of 
$w_{ij}$ in Equation (\ref{eq_sistema_pesi}). \\

\begin{figure}[H]
    \caption{Communities of services}
    \centering
        \includegraphics[width=1\linewidth]{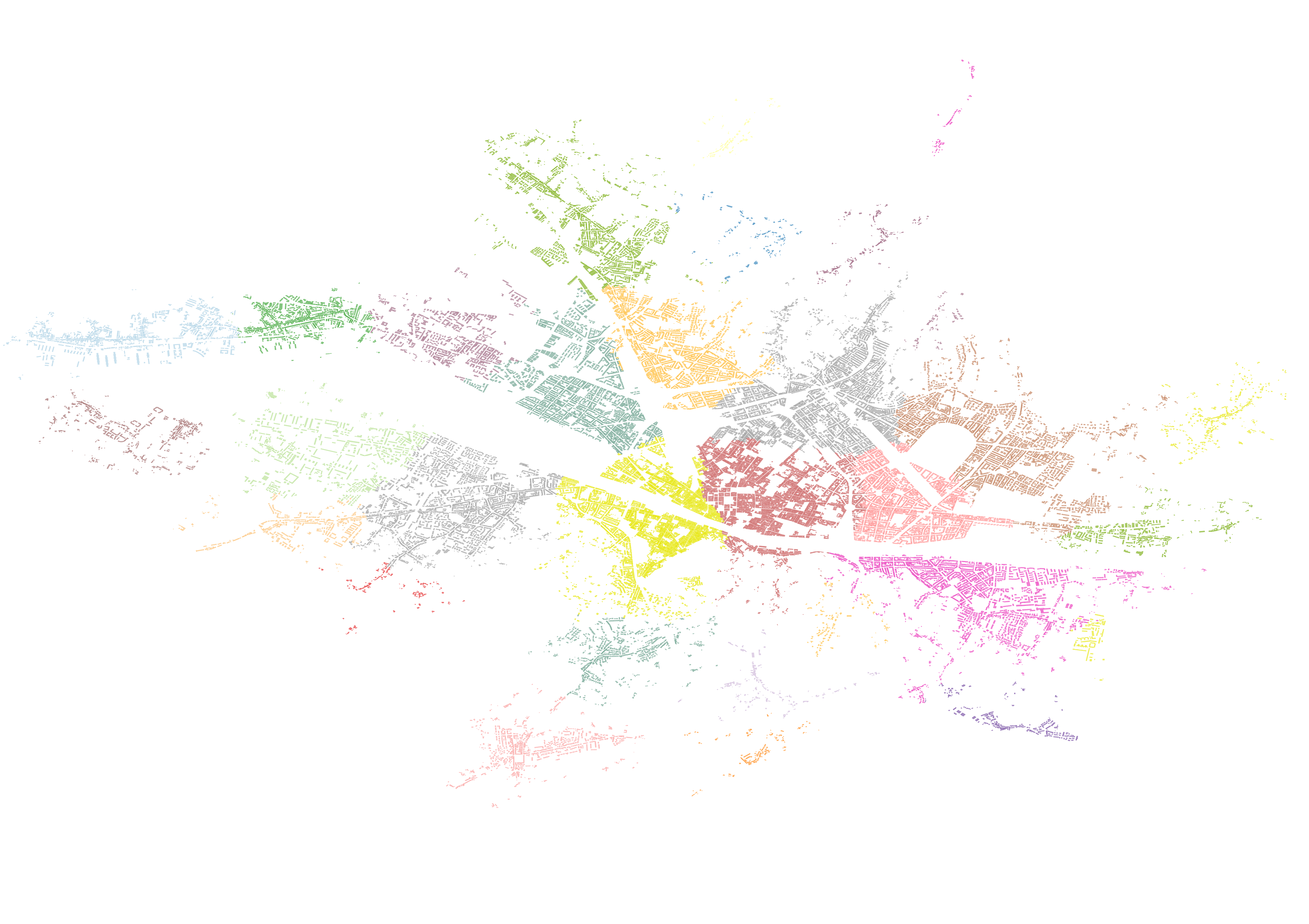} 
    \begin{tablenotes}
      \scriptsize 
      \item Notes: The map above shows what the procedure of communities identification yields in the case of Florence. Colors uniquely identify communities.  
\end{tablenotes}
\label{fig: buildings_comm_map}
\end{figure}

Figure \ref{fig: buildings_comm_map} presents an application of the building community identification to the city of Florence.

\section{Functional redundancy}
\label{sec: redund}
This section expands the previous analysis by incorporating an additional dimension. A building is considered to have access to a service type if at least one POI of that type falls within its isochrone. We do not distinguish between cases with multiple services of the same type and those with only one.
We then introduce the concept of functional redundancy, which refers to the presence of multiple entities within a system performing similar functions, ensuring that the system maintains functionality even if one entity fails or is lost. This concept is well-established in ecological literature on ecosystem sustainability \citep{mori2013response,ricotta2016measuring}, as well as in studies of urban evolution. In fact, functional redundancy is considered a key dimension of urban resilience, defined as the capacity of cities to absorb, adapt to, and recover from various shocks and stresses while maintaining essential functions, structures, and improving sustainability and quality of life for residents \citep{ahern2011fail,elmqvist2019sustainability}.

From an operational perspective, we aim to calculate a functional redundancy index for each building. We assume that every service, regardless of type, has a positive probability of closing or relocating. Our indicator measures the number of closures needed for a building to lose access to a service type, leading to a reduction in the 10-min index. We note that this analysis considers only Points of Interest, excluding green areas. To determine the contribution of each service type to the 10-minute accessibility indicator, mathematical notation must be introduced. The services are divided into $n$ categories, each of which is further partitioned into a finite number of subsets, $p_j$, referred to as subcategories. Note that $p_j$ may be equal to one. Each subcategory, in turn, is divided into a finite number of sets, $m_{j,h}$, representing different service types. Thus, every individual service belongs to a specific type, which is part of a subcategory, which in turn belongs to a broader category. Formally, the entire structure can be expressed using nested set relations. The set $S$ is defined as the union of all categories, i.e.,
\begin{equation*}
    S = \bigcup_{j=1}^{n} C_j
\end{equation*}
where each category $C_j$ is further decomposed into subcategories as follows:
\begin{equation*}
    C_j = \bigcup_{h=1}^{p_j} SC_{j,h}
\end{equation*}
Each subcategory $SC_{j,h}$ is then partitioned into different service types:
\begin{equation*}
    SC_{j,h} = \bigcup_{i=1}^{m_{j,h}} T_{j,h,i}
\end{equation*}
Where $T_{j,h,i}$ is the set of services of type $t_{i}$, belonging to subcategory $SC_{j,h}$, and category $C_j$. Thus, the contribution of each service type to the \textit{10-minute} accessibility indicator is:
\begin{equation*}
     K_{j,h}=\frac{1}{n \cdot p_j \cdot m_{j,h}}
\end{equation*}
For each building $ b^\alpha $, we identify the services $ S^\alpha $ located within its isochrone. Let $T^{\alpha}_{j,h,i}$ be the intersection between $T_{j,h,i}$ and $ S^\alpha $. Thus, the contribution of service type $t_i$ to the indicator obtained from building $b^{\alpha}$ is:
\begin{equation*}
K^{\alpha}_{j,h,i}=
    \begin{cases}
        K_{j,h}/I^{\alpha} & \text{if  } T^{\alpha}_{j,h,i} \neq \emptyset \\
        0 & \text{if  } T^{\alpha}_{j,h,i} = \emptyset
    \end{cases}
\end{equation*}
The functional redundancy indicator for a building is obtained using the following formula:
\begin{equation*}
    R^\alpha = \sum_{j=1}^{n}\sum_{h=1}^{p_j}\sum_{i=1}^{m_{j,h}} \frac{K^{\alpha}_{j,h,i}}{|T^{\alpha}_{j,h,i}|}  
\end{equation*}
Low $R^\alpha$ values indicate low sensitivity to service closure or relocation, implying high functional redundancy. Conversely, high $R^\alpha$ means that even a few service closures or relocations strongly impact the accessibility indicator, $I^\alpha$, indicating low functional redundancy.


\subsection{An application to Florence}

\begin{figure}[H]
    \caption{Exclusive services by typology}
    \centering
        \includegraphics[width=\linewidth]{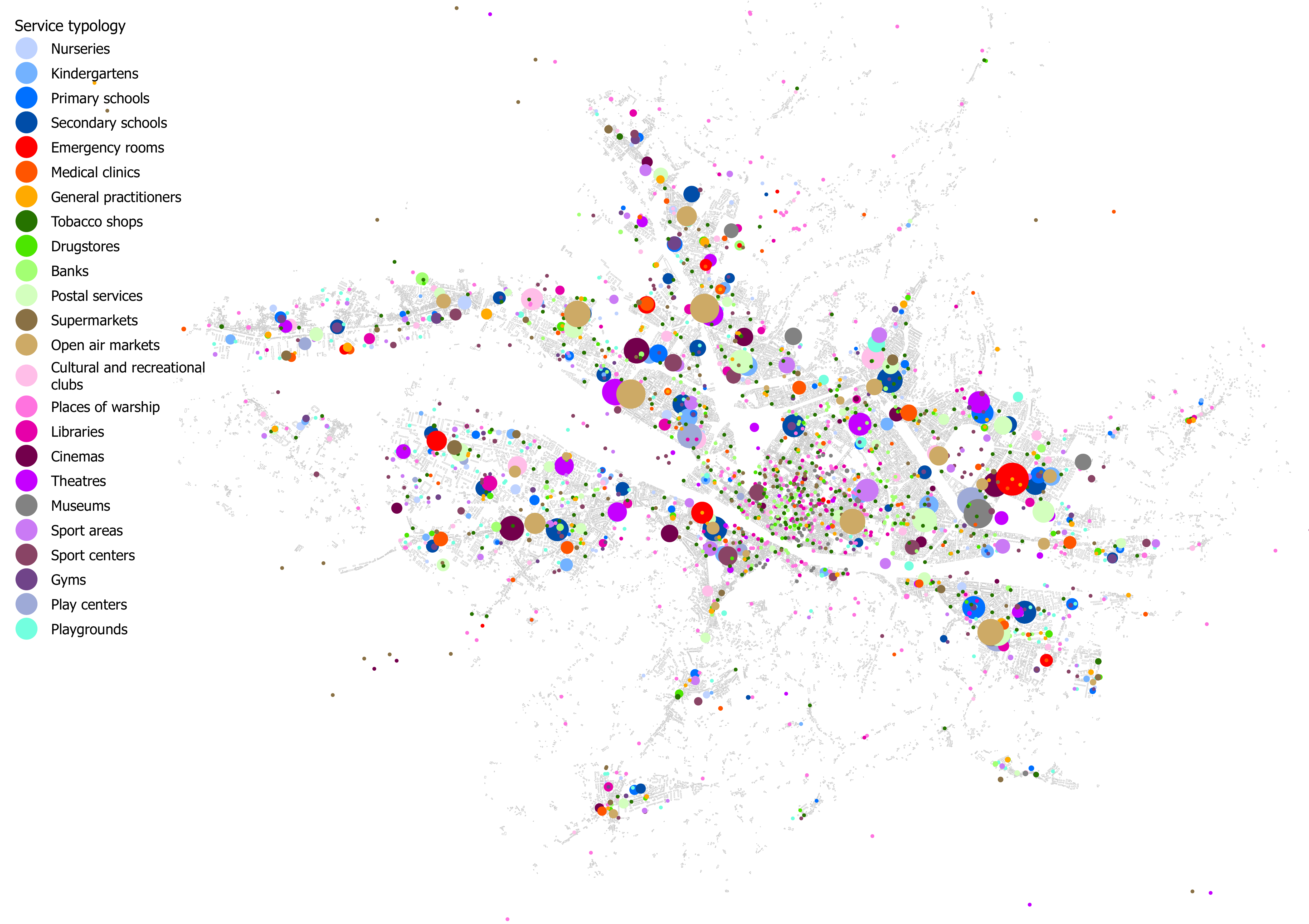} 
    \begin{tablenotes}
      \scriptsize 
      \item Notes: The figure above shows the distribution of exclusive services across the city of Florence. Here, exclusive refers to services accessed by individuals without competition from alternative providers. Different colors indicate distinct service types, while the size of each point corresponds to the number of people served exclusively by that service. Larger points represent a greater population served exclusively by that service, whereas smaller points suggest the opposite. Small points may result from a higher local concentration of similar services, offering multiple substitutes to residents, or from a low number of people being served. The size scale ranges from services reaching as few as 0–500 individuals to those serving between 17,294 and 17,851 people.

\end{tablenotes}
\label{fig: bottle_necks}
\end{figure}

In this section, we apply the previously explained methodology to the city of Florence. We begin by presenting evidence of the significance of functional redundancy and, finally, calculate the $R^\alpha$ index for each building.

As illustrated in Figure \ref{fig: bottle_necks}, we identified the population for which each service is the only provider of its type within a 10-minute walking distance. Larger points represent a greater population served exclusively by that service. Conversely, the presence of many smaller points may indicate that services of the same type coexist within the same isochrones. In such cases, a service shutdown would have a marginal impact on the index value, since the affected population would still have access to at least one other service of the same type. In contrast, the shutdown of a service with a large exclusive user base would significantly affect the index. As shown in Figure \ref{fig: bottle_necks}, the historic center of Florence hosts multiple services of the same type, as highlighted by the presence of many small-sized points.

\begin{figure}[H]
    \caption{10-min accessibility index vs number of accessible services}
    \centering
        \includegraphics[width=1\linewidth]{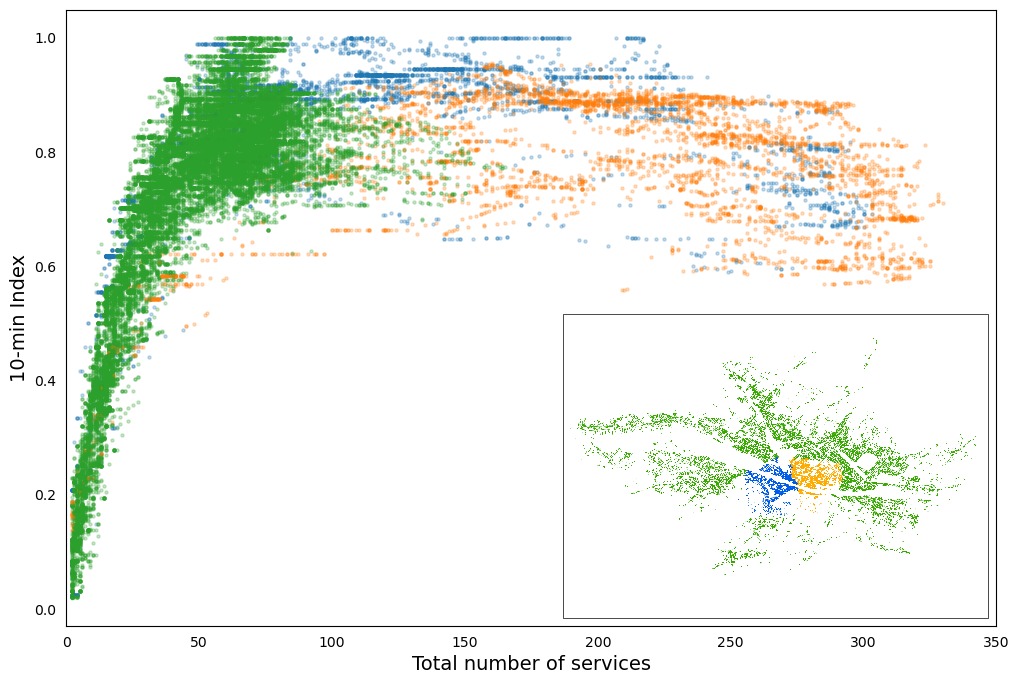} 
    \begin{tablenotes}
      \scriptsize 
      \item Notes: The figure above reports, for each building, the value of the 10-min Index on the vertical axis, and the total number of services located within its isochrone on the horizontal axis. Points in blue and orange represent buildings belonging to the most central communities identified in Section \ref{sec: network} (as usefully indicated in the map in the bottom-left portion of the graph), whereas points in green represent all the other buildings. 
\end{tablenotes}
\label{fig: index_vs_numberofservices}
\end{figure}
\begin{figure}[H]
    \caption{Functional redundancy}
    \centering
        \includegraphics[width=1\linewidth]{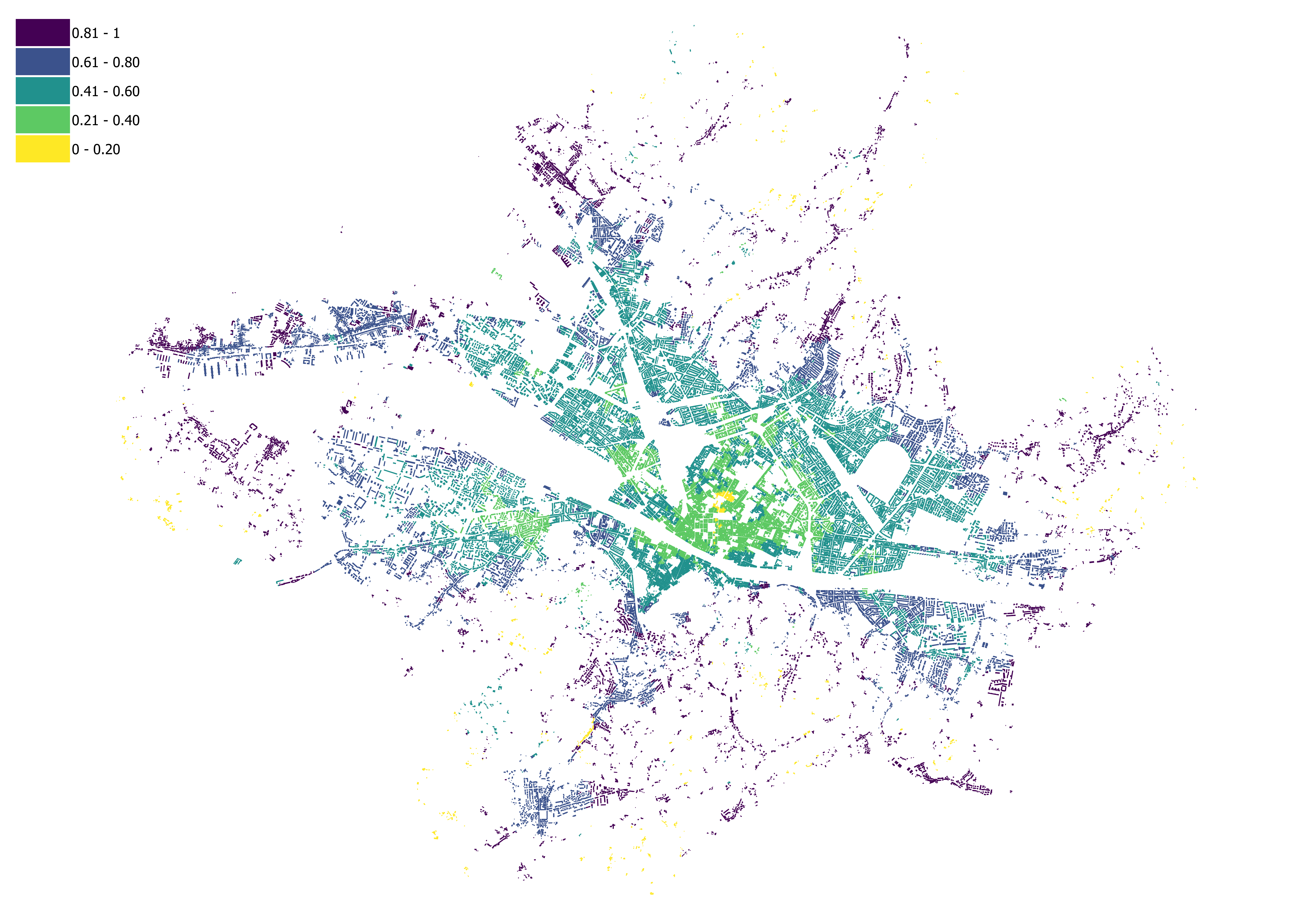} 
    \begin{tablenotes}
      \scriptsize 
      \item Notes: The image above illustrates the spatial distribution of the functional redundancy index across Florence. Lower values indicate higher redundancy, while higher values reflect lower redundancy.
\end{tablenotes}
\label{fig: redundancy}
\end{figure}

The uniqueness of the historic center is also emphasized in Figure \ref{fig: index_vs_numberofservices}, where each point on the scatterplot represents a building $b^\alpha$, measuring the number of services within the isochrone, $|S^\alpha|$, on the horizontal axis and the 10-minute accessibility indicator, $I^\alpha$, on the vertical axis.
As can be seen for the buildings in the two communities of the historic center, in blue and orange, the positive relationship between the number of accessible services and the indicator value is lost. The functional redundancy indicator, $R^\alpha$, for the city of Florence is shown in Figure \ref{fig: redundancy}.
Functional redundancy is high in the historic center, as shown in Figures \ref{fig: bottle_necks} and \ref{fig: index_vs_numberofservices}. Additionally, some peripheral areas in hilly, low-density urban contexts also exhibit high functional redundancy. These areas generally have low $I^\alpha$ values, explained by the presence of multiple services of the same or a few types, making the low $I^\alpha$ value resistant to service closures or relocations.

\section{Discussion and Conclusions}
\label{sec: concl}

In this paper, we introduce a new methodology for quantifying the 10-minute city and present an application to the city of Florence, Italy. The methodological advancements over previous studies lie in the use of residential buildings as the analysis unit and the incorporation of isochrones. Consequently, for each building, we can map the services included within the isochrone centered on that building. Using buildings as the unit of analysis allows for a more granular indicator compared to census tracts or a grid-based approach. Isochrones enable for an accurate consideration of the road network and geographical barriers. Additionally, using building centroids as isochrone origins minimizes the dependency of results on the choice of walking route starting points, which is a major limitation when using census tracts or a grid-based approach.
To construct the index, we consider a given type of service as accessible from a building if at least one facility of that type falls within its isochrone. Service types are then grouped into six categories, each with a corresponding accessibility sub-index. The overall 10-minute accessibility index is calculated as the average of the six sub-indices. Finally, we create an interactive map\footnote{Here is the web address for the interactive map: \url{https://eugeniovicario.github.io/map_10min/}.} that enables users to customize the overall index by adjusting the weights assigned to each category according to their preferences. As a second contribution, we propose a novel methodology to translate accessibility relationships between buildings and services into a weighted and directional network. By applying the InfoMap algorithm, we can identify building communities that share predominant access to the same services. Identifying service-based communities enables the analysis of phenomena such as urban segregation using endogenously generated communities rather than administrative boundaries. Finally, we propose a functional redundancy indicator to assess the resilience of the previously calculated 10-minute accessibility index. This indicator is also calculated at the building level. \\

The proposed methodology, based on the use of isochrones originating from residential buildings, is easily scalable and replicable. Moreover, by using the smallest unit of analysis, i.e. buildings, we avoid comparability issues that may arise when using administrative units or a grid-based approach, which can vary in size and characteristics depending on the context. However, accurately identifying services, and thus points of interest, is a critical step in ensuring the replicability of the methodology. Incomplete data and geolocation inaccuracies heavily affect the results of our calculations, as accessibility measures, given their high spatial resolution, are highly sensitive to even minor spatial variations or missing data. In the absence of comprehensive open-source service maps, we prioritize using data sources that offer full coverage. \\

\noindent Some recurring challenges shape the future development of research on 15-minute (or 10-minute) cities. While widely discussed in urban planning debates, this approach remains largely unsupported by robust empirical evidence. Questions have been raised regarding several aspects, including the challenge of connecting proximity to actual service utilization due to individual preferences and consumption patterns. According to \cite{SU2022103495}, quantifying accessibility of amenities involves two critical issues: the varying relevance of different amenities to consumers and the fact that some amenities, even when geographically distant, can still offer benefits to consumers. The latter limitation creates the need of demonstrating a clear time-use relationship in consumer behavior. Moreover, concerns persist about potential adverse effects, such as urban segregation, and the model's applicability beyond the European context \citep{birkenfeld2023living}. Thus, further research is needed to determine, on the one hand, the actual accessibility of services and, on the other, the extent to which these services are essential. One promising method to tackle these issues is the use of mobility datasets to assess the actual utilization of services. For example, \cite{abbiasov202415} finds correlation between use and time accessibility by leveraging phone mobility data, providing empirical evidence that add a building block to the 15-min theory. Other examples of interesting use of mobility data on related topics are \cite{nilforoshan2023human} and \cite{xu2025using}. Administering surveys to gather insights into citizens’ perceived accessibility and mobility behaviors offers another valuable avenue for extending this research.\\

This methodology can be used to study the evolution of an urban context over time, evaluating the impact of urban planning changes and the reallocation of essential services. \\



\newpage
\setlength\bibsep{0.5pt}
\bibliographystyle{elsarticle-harv}

\begin{thebibliography}{35}
\expandafter\ifx\csname natexlab\endcsname\relax\def\natexlab#1{#1}\fi
\providecommand{\url}[1]{\texttt{#1}}
\providecommand{\href}[2]{#2}
\providecommand{\path}[1]{#1}
\providecommand{\DOIprefix}{doi:}
\providecommand{\ArXivprefix}{arXiv:}
\providecommand{\URLprefix}{URL: }
\providecommand{\Pubmedprefix}{pmid:}
\providecommand{\doi}[1]{\href{http://dx.doi.org/#1}{\path{#1}}}
\providecommand{\Pubmed}[1]{\href{pmid:#1}{\path{#1}}}
\providecommand{\bibinfo}[2]{#2}
\ifx\xfnm\relax \def\xfnm[#1]{\unskip,\space#1}\fi
\bibitem[{Abbiasov et~al.(2024)Abbiasov, Heine, Sabouri, Salazar-Miranda, Santi, Glaeser and Ratti}]{abbiasov202415}
\bibinfo{author}{Abbiasov, T.}, \bibinfo{author}{Heine, C.}, \bibinfo{author}{Sabouri, S.}, \bibinfo{author}{Salazar-Miranda, A.}, \bibinfo{author}{Santi, P.}, \bibinfo{author}{Glaeser, E.}, \bibinfo{author}{Ratti, C.}, \bibinfo{year}{2024}.
\newblock \bibinfo{title}{The 15-minute city quantified using human mobility data}.
\newblock \bibinfo{journal}{Nature Human Behaviour} \bibinfo{volume}{8}, \bibinfo{pages}{445–455}.
\newblock \DOIprefix\doi{https://doi.org/10.1038/s41562-023-01770-y}.
\bibitem[{Ahern(2011)}]{ahern2011fail}
\bibinfo{author}{Ahern, J.}, \bibinfo{year}{2011}.
\newblock \bibinfo{title}{From fail-safe to safe-to-fail: Sustainability and resilience in the new urban world}.
\newblock \bibinfo{journal}{Landscape and urban Planning} \bibinfo{volume}{100}, \bibinfo{pages}{341--343}.
\bibitem[{Akrami et~al.(2024)Akrami, Sliwa and Rynning}]{akrami2024walk}
\bibinfo{author}{Akrami, M.}, \bibinfo{author}{Sliwa, M.W.}, \bibinfo{author}{Rynning, M.K.}, \bibinfo{year}{2024}.
\newblock \bibinfo{title}{Walk further and access more! exploring the 15-minute city concept in oslo, norway}.
\newblock \bibinfo{journal}{Journal of Urban Mobility} \bibinfo{volume}{5}, \bibinfo{pages}{100077}.
\bibitem[{Barrington-Leigh and Millard-Ball(2017)}]{barrington2017world}
\bibinfo{author}{Barrington-Leigh, C.}, \bibinfo{author}{Millard-Ball, A.}, \bibinfo{year}{2017}.
\newblock \bibinfo{title}{The world’s user-generated road map is more than 80\% complete}.
\newblock \bibinfo{journal}{PloS one} \bibinfo{volume}{12}, \bibinfo{pages}{e0180698}.
\bibitem[{Bayer and McMillan(2012)}]{bayer2012tiebout}
\bibinfo{author}{Bayer, P.}, \bibinfo{author}{McMillan, R.}, \bibinfo{year}{2012}.
\newblock \bibinfo{title}{Tiebout sorting and neighborhood stratification}.
\newblock \bibinfo{journal}{Journal of Public Economics} \bibinfo{volume}{96}, \bibinfo{pages}{1129--1143}.
\bibitem[{Billings and Johnson(2016)}]{BILLINGS201613}
\bibinfo{author}{Billings, S.B.}, \bibinfo{author}{Johnson, E.B.}, \bibinfo{year}{2016}.
\newblock \bibinfo{title}{Agglomeration within an urban area}.
\newblock \bibinfo{journal}{Journal of Urban Economics} \bibinfo{volume}{91}, \bibinfo{pages}{13--25}.
\newblock \URLprefix \url{https://www.sciencedirect.com/science/article/pii/S0094119015000704}, \DOIprefix\doi{https://doi.org/10.1016/j.jue.2015.11.002}.
\bibitem[{Birkenfeld et~al.(2023)Birkenfeld, Victoriano-Habit, Alousi-Jones, Soliz and El-Geneidy}]{birkenfeld2023living}
\bibinfo{author}{Birkenfeld, C.}, \bibinfo{author}{Victoriano-Habit, R.}, \bibinfo{author}{Alousi-Jones, M.}, \bibinfo{author}{Soliz, A.}, \bibinfo{author}{El-Geneidy, A.}, \bibinfo{year}{2023}.
\newblock \bibinfo{title}{Who is living a local lifestyle? towards a better understanding of the 15-minute-city and 30-minute-city concepts from a behavioural perspective in montr{\'e}al, canada}.
\newblock \bibinfo{journal}{Journal of Urban Mobility} \bibinfo{volume}{3}, \bibinfo{pages}{100048}.
\bibitem[{Bruno et~al.(2024)Bruno, Monteiro~Melo, Campanelli and Loreto}]{bruno2024universal}
\bibinfo{author}{Bruno, M.}, \bibinfo{author}{Monteiro~Melo, H.P.}, \bibinfo{author}{Campanelli, B.}, \bibinfo{author}{Loreto, V.}, \bibinfo{year}{2024}.
\newblock \bibinfo{title}{A universal framework for inclusive 15-minute cities}.
\newblock \bibinfo{journal}{Nature Cities} , \bibinfo{pages}{1--9}.
\bibitem[{Coombes and Openshaw(1982)}]{coombes1982use}
\bibinfo{author}{Coombes, M.G.}, \bibinfo{author}{Openshaw, S.}, \bibinfo{year}{1982}.
\newblock \bibinfo{title}{The use and definition of travel-to-work areas in great britain: some comments}.
\newblock \bibinfo{journal}{Regional Studies} .
\bibitem[{De~Bellefon et~al.(2021)De~Bellefon, Combes, Duranton, Gobillon and Gorin}]{de2021delineating}
\bibinfo{author}{De~Bellefon, M.P.}, \bibinfo{author}{Combes, P.P.}, \bibinfo{author}{Duranton, G.}, \bibinfo{author}{Gobillon, L.}, \bibinfo{author}{Gorin, C.}, \bibinfo{year}{2021}.
\newblock \bibinfo{title}{Delineating urban areas using building density}.
\newblock \bibinfo{journal}{Journal of Urban Economics} \bibinfo{volume}{125}, \bibinfo{pages}{103226}.
\bibitem[{Elmqvist et~al.(2019)Elmqvist, Andersson, Frantzeskaki, McPhearson, Olsson, Gaffney, Takeuchi and Folke}]{elmqvist2019sustainability}
\bibinfo{author}{Elmqvist, T.}, \bibinfo{author}{Andersson, E.}, \bibinfo{author}{Frantzeskaki, N.}, \bibinfo{author}{McPhearson, T.}, \bibinfo{author}{Olsson, P.}, \bibinfo{author}{Gaffney, O.}, \bibinfo{author}{Takeuchi, K.}, \bibinfo{author}{Folke, C.}, \bibinfo{year}{2019}.
\newblock \bibinfo{title}{Sustainability and resilience for transformation in the urban century}.
\newblock \bibinfo{journal}{Nature sustainability} \bibinfo{volume}{2}, \bibinfo{pages}{267--273}.
\bibitem[{Epple and Platt(1998)}]{epple1998equilibrium}
\bibinfo{author}{Epple, D.}, \bibinfo{author}{Platt, G.J.}, \bibinfo{year}{1998}.
\newblock \bibinfo{title}{Equilibrium and local redistribution in an urban economy when households differ in both preferences and incomes}.
\newblock \bibinfo{journal}{Journal of urban Economics} \bibinfo{volume}{43}, \bibinfo{pages}{23--51}.
\bibitem[{Epple and Zelenitz(1981)}]{epple1981implications}
\bibinfo{author}{Epple, D.}, \bibinfo{author}{Zelenitz, A.}, \bibinfo{year}{1981}.
\newblock \bibinfo{title}{The implications of competition among jurisdictions: Does tiebout need politics?}
\newblock \bibinfo{journal}{Journal of Political Economy} \bibinfo{volume}{89}, \bibinfo{pages}{1197--1217}.
\bibitem[{{European Environment Agency}(2006)}]{EEA2006}
\bibinfo{author}{{European Environment Agency}}, \bibinfo{year}{2006}.
\newblock \bibinfo{title}{EEA Environmental Statement 2006}.
\newblock \bibinfo{type}{Corporate Document 1/2006}. European Environment Agency.
\newblock \URLprefix \url{https://www.eea.europa.eu/en/analysis/publications/report_2006_0707_150910}.
\bibitem[{Glaeser and Kahn(2004)}]{glaeser2004sprawl}
\bibinfo{author}{Glaeser, E.L.}, \bibinfo{author}{Kahn, M.E.}, \bibinfo{year}{2004}.
\newblock \bibinfo{title}{Sprawl and urban growth}, in: \bibinfo{booktitle}{Handbook of regional and urban economics}. \bibinfo{publisher}{Elsevier}. volume~\bibinfo{volume}{4}, pp. \bibinfo{pages}{2481--2527}.
\bibitem[{Glaeser and Kahn(2010)}]{glaeser2010greenness}
\bibinfo{author}{Glaeser, E.L.}, \bibinfo{author}{Kahn, M.E.}, \bibinfo{year}{2010}.
\newblock \bibinfo{title}{The greenness of cities: Carbon dioxide emissions and urban development}.
\newblock \bibinfo{journal}{Journal of urban economics} \bibinfo{volume}{67}, \bibinfo{pages}{404--418}.
\bibitem[{ISTAT(2017)}]{istat2017urban}
\bibinfo{author}{ISTAT}, \bibinfo{year}{2017}.
\newblock \bibinfo{title}{Forme, livelli e dinamiche dell'urbanizzazione}.
\newblock \URLprefix \url{https://www.istat.it/it/files/2017/05/Urbanizzazione.pdf}.
\bibitem[{Lancichinetti and Fortunato(2009)}]{lancichinetti2009community}
\bibinfo{author}{Lancichinetti, A.}, \bibinfo{author}{Fortunato, S.}, \bibinfo{year}{2009}.
\newblock \bibinfo{title}{Community detection algorithms: a comparative analysis}.
\newblock \bibinfo{journal}{Physical Review E—Statistical, Nonlinear, and Soft Matter Physics} \bibinfo{volume}{80}, \bibinfo{pages}{056117}.
\bibitem[{Moreno et~al.(2021)Moreno, Allam, Chabaud, Gall and Pratlong}]{moreno2021introducing}
\bibinfo{author}{Moreno, C.}, \bibinfo{author}{Allam, Z.}, \bibinfo{author}{Chabaud, D.}, \bibinfo{author}{Gall, C.}, \bibinfo{author}{Pratlong, F.}, \bibinfo{year}{2021}.
\newblock \bibinfo{title}{Introducing the “15-minute city”: Sustainability, resilience and place identity in future post-pandemic cities}.
\newblock \bibinfo{journal}{Smart cities} \bibinfo{volume}{4}, \bibinfo{pages}{93--111}.
\bibitem[{Mori et~al.(2013)Mori, Furukawa and Sasaki}]{mori2013response}
\bibinfo{author}{Mori, A.S.}, \bibinfo{author}{Furukawa, T.}, \bibinfo{author}{Sasaki, T.}, \bibinfo{year}{2013}.
\newblock \bibinfo{title}{Response diversity determines the resilience of ecosystems to environmental change}.
\newblock \bibinfo{journal}{Biological reviews} \bibinfo{volume}{88}, \bibinfo{pages}{349--364}.
\bibitem[{Murgante et~al.(2024)Murgante, Valluzzi and Annunziata}]{murgante2024developing}
\bibinfo{author}{Murgante, B.}, \bibinfo{author}{Valluzzi, R.}, \bibinfo{author}{Annunziata, A.}, \bibinfo{year}{2024}.
\newblock \bibinfo{title}{Developing a 15-minute city: Evaluating urban quality using configurational analysis. the case study of terni and matera, italy}.
\newblock \bibinfo{journal}{Applied Geography} \bibinfo{volume}{162}, \bibinfo{pages}{103171}.
\bibitem[{Nilforoshan et~al.(2023)Nilforoshan, Looi, Pierson, Villanueva, Fishman, Chen, Sholar, Redbird, Grusky and Leskovec}]{nilforoshan2023human}
\bibinfo{author}{Nilforoshan, H.}, \bibinfo{author}{Looi, W.}, \bibinfo{author}{Pierson, E.}, \bibinfo{author}{Villanueva, B.}, \bibinfo{author}{Fishman, N.}, \bibinfo{author}{Chen, Y.}, \bibinfo{author}{Sholar, J.}, \bibinfo{author}{Redbird, B.}, \bibinfo{author}{Grusky, D.}, \bibinfo{author}{Leskovec, J.}, \bibinfo{year}{2023}.
\newblock \bibinfo{title}{Human mobility networks reveal increased segregation in large cities}.
\newblock \bibinfo{journal}{Nature} \bibinfo{volume}{624}, \bibinfo{pages}{586--592}.
\bibitem[{OECD(2012)}]{oecd2012compact}
\bibinfo{author}{OECD}, \bibinfo{year}{2012}.
\newblock \bibinfo{title}{Compact City Policies: A Comparative Assessment}.
\newblock OECD Green Growth Studies, \bibinfo{publisher}{OECD Publishing}, \bibinfo{address}{Paris}.
\newblock \URLprefix \url{https://doi.org/10.1787/9789264167865-en}, \DOIprefix\doi{10.1787/9789264167865-en}.
\bibitem[{OECD(2024)}]{oecd2022life}
\bibinfo{author}{OECD}, \bibinfo{year}{2024}.
\newblock \bibinfo{title}{How's Life? 2024: Well-being and Resilience in Times of Crisis}.
\newblock \bibinfo{publisher}{OECD Publishing}, \bibinfo{address}{Paris}.
\newblock \DOIprefix\doi{https://doi.org/10.1787/90ba854a-en}.
\bibitem[{OECD and {European Commission}(2020)}]{oecd2020cities}
\bibinfo{author}{OECD}, \bibinfo{author}{{European Commission}}, \bibinfo{year}{2020}.
\newblock \bibinfo{title}{Cities in the World: A New Perspective on Urbanisation}.
\newblock OECD Urban Studies, \bibinfo{publisher}{OECD Publishing}, \bibinfo{address}{Paris}.
\newblock \DOIprefix\doi{https://doi.org/10.1787/d0efcbda-en}.
\bibitem[{Olivari et~al.(2023)Olivari, Cipriano, Napolitano and Giovannini}]{olivari2023italian}
\bibinfo{author}{Olivari, B.}, \bibinfo{author}{Cipriano, P.}, \bibinfo{author}{Napolitano, M.}, \bibinfo{author}{Giovannini, L.}, \bibinfo{year}{2023}.
\newblock \bibinfo{title}{Are italian cities already 15-minute? presenting the next proximity index: A novel and scalable way to measure it, based on open data}.
\newblock \bibinfo{journal}{Journal of Urban Mobility} \bibinfo{volume}{4}, \bibinfo{pages}{100057}.
\bibitem[{Rappaport(2008)}]{rappaport2008consumption}
\bibinfo{author}{Rappaport, J.}, \bibinfo{year}{2008}.
\newblock \bibinfo{title}{Consumption amenities and city population density}.
\newblock \bibinfo{journal}{Regional science and urban economics} \bibinfo{volume}{38}, \bibinfo{pages}{533--552}.
\bibitem[{Ricotta et~al.(2016)Ricotta, de~Bello, Moretti, Caccianiga, Cerabolini and Pavoine}]{ricotta2016measuring}
\bibinfo{author}{Ricotta, C.}, \bibinfo{author}{de~Bello, F.}, \bibinfo{author}{Moretti, M.}, \bibinfo{author}{Caccianiga, M.}, \bibinfo{author}{Cerabolini, B.E.}, \bibinfo{author}{Pavoine, S.}, \bibinfo{year}{2016}.
\newblock \bibinfo{title}{Measuring the functional redundancy of biological communities: a quantitative guide}.
\newblock \bibinfo{journal}{Methods in Ecology and Evolution} \bibinfo{volume}{7}, \bibinfo{pages}{1386--1395}.
\bibitem[{Roback(1982)}]{roback1982wages}
\bibinfo{author}{Roback, J.}, \bibinfo{year}{1982}.
\newblock \bibinfo{title}{Wages, rents, and the quality of life}.
\newblock \bibinfo{journal}{Journal of political Economy} \bibinfo{volume}{90}, \bibinfo{pages}{1257--1278}.
\bibitem[{Rosenthal et~al.(2022)Rosenthal, Strange and Urrego}]{ROSENTHAL2022103381}
\bibinfo{author}{Rosenthal, S.S.}, \bibinfo{author}{Strange, W.C.}, \bibinfo{author}{Urrego, J.A.}, \bibinfo{year}{2022}.
\newblock \bibinfo{title}{Jue insight: Are city centers losing their appeal? commercial real estate, urban spatial structure, and covid-19}.
\newblock \bibinfo{journal}{Journal of Urban Economics} \bibinfo{volume}{127}, \bibinfo{pages}{103381}.
\newblock \DOIprefix\doi{https://doi.org/10.1016/j.jue.2021.103381}. \bibinfo{note}{jUE Insights: COVID-19 and Cities}.
\bibitem[{Rosvall and Bergstrom(2008)}]{rosvall2008maps}
\bibinfo{author}{Rosvall, M.}, \bibinfo{author}{Bergstrom, C.T.}, \bibinfo{year}{2008}.
\newblock \bibinfo{title}{Maps of random walks on complex networks reveal community structure}.
\newblock \bibinfo{journal}{Proceedings of the national academy of sciences} \bibinfo{volume}{105}, \bibinfo{pages}{1118--1123}.
\bibitem[{Staricco(2022)}]{staricco202215}
\bibinfo{author}{Staricco, L.}, \bibinfo{year}{2022}.
\newblock \bibinfo{title}{15-, 10-or 5-minute city? a focus on accessibility to services in turin, italy}.
\newblock \bibinfo{journal}{Journal of urban mobility} \bibinfo{volume}{2}, \bibinfo{pages}{100030}.
\bibitem[{Su(2022)}]{SU2022103495}
\bibinfo{author}{Su, Y.}, \bibinfo{year}{2022}.
\newblock \bibinfo{title}{Measuring the value of urban consumption amenities: A time-use approach}.
\newblock \bibinfo{journal}{Journal of Urban Economics} \bibinfo{volume}{132}, \bibinfo{pages}{103495}.
\newblock \URLprefix \url{https://www.sciencedirect.com/science/article/pii/S0094119022000729}, \DOIprefix\doi{https://doi.org/10.1016/j.jue.2022.103495}.
\bibitem[{WHO(2017)}]{europe2017urban}
\bibinfo{author}{WHO}, \bibinfo{year}{2017}.
\newblock \bibinfo{title}{Urban green spaces: a brief for action}.
\newblock \bibinfo{journal}{World Health Organization. Regional Office for Europe} .
\bibitem[{Xu et~al.(2025)Xu, Wang, Moro, Chen, Salazar~Miranda, Gonz{\'a}lez, Tizzoni, Song, Ratti, Bettencourt et~al.}]{xu2025using}
\bibinfo{author}{Xu, F.}, \bibinfo{author}{Wang, Q.}, \bibinfo{author}{Moro, E.}, \bibinfo{author}{Chen, L.}, \bibinfo{author}{Salazar~Miranda, A.}, \bibinfo{author}{Gonz{\'a}lez, M.C.}, \bibinfo{author}{Tizzoni, M.}, \bibinfo{author}{Song, C.}, \bibinfo{author}{Ratti, C.}, \bibinfo{author}{Bettencourt, L.}, et~al., \bibinfo{year}{2025}.
\newblock \bibinfo{title}{Using human mobility data to quantify experienced urban inequalities}.
\newblock \bibinfo{journal}{Nature Human Behaviour} , \bibinfo{pages}{1--11}.

\end{thebibliography}




\end{document}